\documentclass[ALICE,manyauthors]{cernphprep}
\pdfoutput=1

\usepackage{rotating}
\usepackage[normalem]{ulem}
\usepackage{amssymb,amsmath}
\usepackage{graphicx}
\usepackage{subfigure}
\usepackage{amssymb}
\usepackage{epstopdf}
\usepackage{cite}
\usepackage{multirow}
\usepackage{color}
\usepackage{amsmath,bm}
\usepackage{amsbsy}

\usepackage{hyperref}

\usepackage{datetime}
\usdate
\usepackage{lineno}
\usepackage[toc,page]{appendix}
\usepackage[section] {placeins}

\newcommand{\CommentBlock}[1]{}

\newcommand{\REA}{\ensuremath{R_\mathrm{EA}}}

\newcommand{\dAu}{d--Au}
\newcommand{\AuAu}{Au--Au}
\newcommand{\PbPb}{Pb--Pb}
\newcommand{\pPb}{p--Pb}

\newcommand{\pp}{pp}
\newcommand{\pA}{p--A}
\newcommand{\aaa}{\ensuremath{A+A}}
\newcommand{\sqrtsNN}{\ensuremath{\sqrt{s_{\rm NN}}}}
\newcommand{\sqrts}{\ensuremath{\sqrt{s}}}
\newcommand{\rr}{\ensuremath{R}}

\newcommand{\gev}{\ensuremath{\mathrm{GeV/}c}}

\newcommand{\TpPb}{\ensuremath{T_{\mathrm{pPb}}}}
\newcommand{\TdAu}{\ensuremath{T_{\mathrm{dAu}}}}

\newcommand{\TaA}{\ensuremath{T_{\mathrm{aA}}}}

\newcommand{\kT}{\ensuremath{k_\mathrm{T}}}
\newcommand{\antikT}{anti-\ensuremath{k_\mathrm{T}}}

\newcommand{\pT}{\ensuremath{p_\mathrm{T}}}

\newcommand{\pTjet}{\ensuremath{p_{\mathrm{T,jet}}}}
\newcommand{\pTjetch}{\ensuremath{p_\mathrm{T,jet}^\mathrm{ch}}}

\newcommand{\pTtrig}{\ensuremath{p_{\mathrm{T,trig}}}}

\newcommand{\etajet}{\ensuremath{\eta_\mathrm{jet}}}

\newcommand{\dpT}{\ensuremath{\delta{\pT}}}

\newcommand{\pTembed}{\ensuremath{\pT^{\rm{embed}}}}

\newcommand{\qhat}{\ensuremath{\hat{q}}}

\newcommand{\NtrigX}{\ensuremath{N_{\rm{trig}}}}

\newcommand{\Njets}{\ensuremath{N_{\rm{jets}}}}

\newcommand{\RCP}{\ensuremath{R_\mathrm{CP}}}

\newcommand{\RCPstar}{\ensuremath{R_\mathrm{CP}^{*}}}

\newcommand{\Drecoil}{\ensuremath{\Delta_\mathrm{recoil}}}
\newcommand{\cRef}{\ensuremath{c_\mathrm{Ref}}}

\newcommand{\DrecoilM}{\ensuremath{\Delta^\mathrm{M}_\mathrm{recoil}}}
\newcommand{\DrecoilT}{\ensuremath{\Delta^\mathrm{T}_\mathrm{recoil}}}

\newcommand{\pTraw}{\ensuremath{p_\mathrm{T,jet}^\mathrm{raw,ch}}}

\newcommand{\pTreco}{\ensuremath{p_\mathrm{T,jet}^\mathrm{reco,ch}}}

\newcommand{\Ajet}{\ensuremath{A_\mathrm{jet}}}

\newcommand{\pTdet}{\ensuremath{p_\mathrm{T,jet}^\mathrm{det}}}

\newcommand{\pTpart}{\ensuremath{p_\mathrm{T,jet}^\mathrm{part}}}

\newcommand{\dphi}{\ensuremath{\Delta\varphi}}

\newcommand{\TTSig}{\ensuremath{\mathrm{TT}_{\mathrm{Sig}}}}
\newcommand{\TTRef}{\ensuremath{\mathrm{TT}_{\mathrm{Ref}}}}

\newcommand{\Rbkgd}{\ensuremath{R_{\mathrm{bkgd}}}}
\newcommand{\Rinstr}{\ensuremath{R_{\mathrm{instr}}}}
\newcommand{\Rfull}{\ensuremath{R_{\mathrm{full}}}}

\newcommand{\pTgen}{\ensuremath{p_{\mathrm{T,jet}}^{\mathrm{part}}}}

\newcommand{\yLAB}{\ensuremath{y_\mathrm{LAB}}}
\newcommand{\yNN}{\ensuremath{y_\mathrm{NN}}}
\newcommand{\ystar}{\ensuremath{y^*}}
\newcommand{\ystarTT}{\ensuremath{y^*_\mathrm{TT}}}
\newcommand{\ystarjet}{\ensuremath{y^*_\mathrm{jet}}}

\begin{document}

\begin{titlepage}
\PHyear{2017}
\PHnumber{324}          
\PHdate{10 December}              

\title{Constraints on jet quenching in \pPb\ collisions at $\boldsymbol{\sqrt{s_{\rm NN}}} = 5.02$\,TeV
measured by the event-activity dependence of semi-inclusive hadron-jet distributions}
\ShortTitle{Constraints on jet quenching in \pPb\ collisions} 

\Collaboration{ALICE Collaboration%
         \thanks{See Appendix~\ref*{app:collab} for the list of collaboration
                      members}}
\ShortAuthor{ALICE Collaboration}      

\begin{abstract}
The ALICE Collaboration reports the measurement of semi-inclusive distributions 
of charged-particle jets recoiling from a high-transverse momentum trigger 
hadron in \pPb\ collisions at $\sqrtsNN =5.02$\,TeV. Jets are reconstructed from charged-particle tracks 
using the \antikT\ algorithm 
with resolution parameter $\rr=0.2$ and 0.4. A data-driven statistical approach 
is used to correct the uncorrelated background jet yield. Recoil jet distributions 
are reported for jet transverse momentum $15<\pTjetch\ <50$\,\gev\ and are 
compared in various intervals of \pPb\ event activity, based on 
charged-particle multiplicity and zero-degree neutral energy in the forward 
(Pb-going) direction. The semi-inclusive observable is self-normalized and such 
comparisons do not require the interpretation of \pPb\ event activity in terms 
of  collision geometry, in contrast to inclusive jet observables. These 
measurements provide new constraints on the magnitude of jet quenching in small 
systems at the LHC. In \pPb\ collisions with high event activity, the average 
medium-induced out-of-cone energy transport for jets with $\rr=0.4$ and 
$15<\pTjetch<50$\, \gev\ is measured to be less than 0.4\,\gev\ at 90\% confidence, which is 
over an order of magnitude smaller than a similar measurement for central \PbPb\ 
collisions at $\sqrtsNN=2.76$\,TeV. Comparison is made to theoretical 
calculations of jet quenching in small systems, and to inclusive jet 
measurements in \pPb\ collisions selected by event activity at the LHC and in \dAu\ collisions at RHIC.
\end{abstract}

\maketitle
\end{titlepage}
 

\section{Introduction}
\label{sec:Intro}

The collision of heavy nuclei at high energies generates a Quark-Gluon Plasma 
(QGP), a dense, highly 
inviscid, strongly-coupled fluid governed by sub-nucleonic degrees of 
freedom~\cite{Muller:2012zq}.
While the structure and dynamical behavior of the QGP arise at the microscopic 
level from the interactions between quarks and gluons that are described by 
Quantum Chromodynamics (QCD), the QGP also exhibits 
emergent collective behavior. Current understanding of the properties of the QGP 
is  based primarily on two phenomena observed in high energy nuclear 
collisions and their comparison to theoretical 
calculations: strong collective flow \cite{Heinz:2013th}, and jet quenching, 
which arises from interaction of energetic jets with the medium 
\cite{Burke:2013yra}. 

Jets in hadronic collisions are generated by hard (high 
momentum transfer $Q^2$) interactions between quarks and gluons from the 
projectiles, with outgoing quarks and gluons from the interaction observed in 
detectors as 
correlated sprays of hadrons (``jets"). Theoretical 
calculations of jet production based on perturbative QCD (pQCD) are in excellent 
agreement over a broad kinematic range with  
jet measurements in \pp\ collisions at the Large Hadron Collider 
(LHC)~\cite{Abelev:2013fn,Aad:2014vwa,Khachatryan:2016mlc,Dasgupta:2014yra}. Measurements in 
\pp\ collisions of charged-particle jets, which consist of the charged component 
of the hadronic jet shower, are also well-described by QCD-based Monte Carlo 
calculations
~\cite{ALICE:2014dla,Adam:2015pbpb}.

In nuclear collisions, 
the interaction of jets 
with the QGP is expected to modify the observed rate of jet production and 
internal jet structure. 
Indeed,  
marked effects due to jet 
quenching have been observed for high transverse momentum (high-\pT) hadrons and 
jets in 
central \AuAu\ collisions at the Relativistic Heavy Ion Collider (RHIC)
\cite{Adcox:2001jp,Adare:2012wg,Adare:2010ry,Adare:2012qi,Adler:2002xw,Adler:2002tq,Adams:2003kv,Adams:2006yt,Adamczyk:2013jei,Adamczyk:2016fqm,Adamczyk:2017yhe} 
and in central \PbPb\ collisions 
at the 
LHC~\cite{Aamodt:2011vg,Abelev:2012hxa,Abelev:2013kqa,Adam:2015ewa,Adam:2015pbpb,Adam:2016jp,Aad:2010bu,Aad:2014bxa,CMS:2012aa,Chatrchyan:2012wg,Chatrchyan:2012nia,Chatrchyan:2012gt,Khachatryan:2016jfl}. 
Jets therefore provide well-calibrated probes of the QGP.

Measurements of asymmetric \pPb\ collisions at the LHC 
and of light nucleus--Au collisions at RHIC reveal evidence of collective effects that are similar 
in magnitude to those
observed in symmetric collisions of heavy nuclei 
\cite{Abelev:2013ppb,Abelev:2013ppbId,Abelev:2014mda,Adam:2015bka,Aad:2012gla,Aad:2013fja,Aad:2014lta,Khachatryan:2010gv, 
Chatrchyan:2013nka,Chatrchyan:2013ppb,Khachatryan:2014jra,Khachatryan:2015ppb,Adare:2013piz,Adare:2014keg,Adare:2015ctn,Aidala:2016vgl,Adamczyk:2014fcx,Adamczyk:2015xjc}. 
These measurements in asymmetric systems are reproduced both by model 
calculations 
that incorporate 
a locally thermalized hydrodynamic medium in the final state, and by 
calculations without 
QGP but with large fluctuations in the initial-state wavefunctions of the 
projectiles 
(see \cite{Salgado:2016jws} and references therein).
This raises the question whether a QGP is in fact generated in such light 
asymmetric systems, which were initially thought to be too small for the 
formation of a quasi-equilibrated fireball of matter in the final state 
\cite{Salgado:2016jws}. 
Additional measurements, in particular to explore jet quenching in \pA\ 
collisions, will 
help to resolve this picture and to clarify the 
nature of equilibration in strongly-interacting matter.

There are several theoretical calculations currently available of jet quenching 
effects in \pPb\ collisions at the LHC, which differ in their predictions. 
The calculation in~\cite{Tyw:2014ppb} estimates the size of the region of high 
energy density to have a radius that is a factor 2 smaller in \pPb\ than in central \PbPb\ 
collisions, with jet transport parameter \qhat, assumed to be proportional to 
charged multiplicity, to be a factor 7 smaller. Jet energy loss, which is 
proportional to \qhat\ and depends on path length of the jet in the medium, is consequently expected in this calculation to be much 
smaller in \pPb\ than in central \PbPb\ collisions.
In contrast, a model calculation based on one-dimensional Bjorken hydrodynamics 
predicts large initial energy density in high multiplicity \pp\ and \pPb\ 
collisions~\cite{Zakharov:2013gya}; this energy density corresponds to jet 
energy loss of several GeV, which is similar in magnitude to jet energy loss 
measured in central \PbPb\ 
collisions\cite{Adam:2015pbpb}. A calculation based on pQCD 
at next-to-leading order (NLO) finds negligible jet quenching effects for 
inclusive jet production in \pPb\ collisions at 
$\sqrtsNN=5$\,TeV~\cite{Chen:2015qmd}. 
Finally, a QCD calculation of initial-state energy loss in cold nuclear 
matter (CNM) finds significant suppression of inclusive jet production for 
small-impact parameter \pPb\ collisions at 
$\sqrtsNN=5$\,TeV~\cite{Kang:2015mta}.

Experimental searches for jet quenching effects in \dAu\ collisions at RHIC and 
in \pPb\ collisions at the LHC have been carried out with high-\pT\ hadrons 
and reconstructed jets. These studies utilize both Minimum Bias 
(MB) events and more-differential event selection, in which events are 
characterized in terms of ``event 
activity'' (EA) based on central charged-particle multiplicity (ALICE~\cite{Abelev:2015cnt}); forward charged-particle multiplicity (STAR~\cite{Adams:2003im}, PHENIX~\cite{Adler:2006wg,Adare:2016dau}, ALICE~\cite{Abelev:2015cnt}); forward transverse 
energy (ATLAS~\cite{Aad:2016zif}, CMS~\cite{Chatrchyan:2014hqa}); or zero-degree neutral energy (STAR~\cite{Adams:2003im}, ALICE~\cite{Abelev:2015cnt}); where ``forward" and ``zero-degree" refer to the direction of the heavy nuclear 
projectile. 

Inclusive hadron measurements in \dAu\ collisions at RHIC ~\cite{Adams:2003im,Adler:2006wg} exhibit  yield 
enhancement in the region $2<\pT<5$ \gev, which is commonly attributed to 
multiple scattering in the initial state, with no significant yield modification 
at higher \pT\ and with no significant difference observed between the MB and 
EA-selected distributions. For inclusive hadron measurements in \pPb\ 
collisions at the LHC, ALICE does not observe 
significant  yield modification for $\pT>8$ \gev\ in both MB and 
EA-selected events \cite{Abel:2015ppb,Abelev:2015cnt} while ATLAS and 
CMS observe yield enhancement for \pT\ greater than $\sim30$ \gev\ in 
MB events 
\cite{Aad:2016zif,Khachatryan:2015xaa,Khachatryan:2016odn}, and ATLAS observes additional 
dependence on EA \cite{Aad:2016zif}.

For inclusive jet production, no significant yield modification has been 
observed in MB 
\pPb\ collisions at the LHC and MB \dAu\ 
collisions at RHIC \cite{Adam:2015hoa,Aad:2015ppb,
Adare:2016dau,Khachatryan:2016xdg}. However, 
measurements
by the PHENIX collaboration at RHIC~\cite{Adare:2016dau} and the ATLAS 
collaboration at the LHC~\cite{Aad:2015ppb}  
find apparent enhancement of the inclusive jet yield in EA-selected event 
populations thought to 
be biased towards large 
impact parameter in such asymmetric systems
(``peripheral collisions"), with compensating suppression for event populations
assigned small impact parameter 
(``central collisions"), while the ALICE collaboration finds no such yield 
modification as a function of event ``centrality"~\cite{Adam:2016ppb}.

Measurement of jet quenching effects with 
inclusive processes requires scaling of the inclusive yield from a 
reference collision system (usually \pp) by the nuclear overlap function 
$\left<\TaA\right>$, with the angle brackets $\left<\ldots\right>$ indicating an 
average over the event population; for 
current measurements, ``aA" denotes \dAu\ at RHIC and \pPb\ at the LHC. For an EA-selected population, $\left<\TaA\right>$ is 
calculated by correlating EA with collision geometry and
applying Glauber modeling \cite{Miller:2007ri}. 
However, the correlation of EA with collision geometry in \pPb\ 
collisions is obscured by large fluctuations in the EA observables 
\cite{Abelev:2015cnt}, and can be biased by conservation laws 
and by dynamical correlations when 
measuring high $Q^2$ processes 
\cite{Adare:2013nff,Maj:2014ppb,Perepelitsa:2014yta,Bzdak:2014rca,Armesto:2015kwa,Zakharov:2016zqc}. 
Color fluctuations in the proton wavefunction may induce a bias in 
soft particle production for \pPb\ events tagged by a hard process, thereby 
biasing the correlation between EA and collision 
geometry~\cite{Alvioli:2013vk,Alvioli:2014sba,Alvioli:2014eda,McGlinchey:2016ssj}.
A model calculation shows that selection bias can modify the scaling factor for 
jet production in peripheral \aaa\ relative to \pp\ collisions, generating an
apparent suppression of jet production in peripheral \aaa\ collisions if the 
Glauber calculation does not take this effect into account~\cite{Morsch:2017br}; 
similar considerations 
apply to asymmetric collision systems.

While Glauber modeling for peripheral
\dAu\ collisions at RHIC has been validated experimentally 
for moderate $Q^2$ processes using a proton-stripping process and 
knowledge of the deuteron wavefunction 
\cite{Adams:2003im}, 
no such check is possible with the proton beam at the LHC. It is therefore 
crucial to measure the EA-dependence of jet 
quenching effects in \pPb\ collisions at the LHC with correlation
observables that do not require the interpretation of EA in 
terms of collision geometry.

A correlation measurement of dijet transverse-momentum balance in \pPb\ 
collisions at $\sqrtsNN=5.02$\,TeV finds no significant difference from a simulated 
\pp\ reference distribution, independent of EA~\cite{Chatrchyan:2014hqa}. 
Measurements of dijet acoplanarity, which can be generated by both initial-state 
and final-state effects, likewise find no significant modification due to 
nuclear matter effects in EA-selected \pPb\ collisions at $\sqrtsNN=5.02$\,TeV, 
relative to simulated distributions for \pp\ 
collisions~\cite{Chatrchyan:2014hqa,Adam:2015xea}. While these measurements 
provide qualitative indications, based on comparison to simulations, that final-state 
jet quenching effects in high-EA \pPb\ 
collisions are small, quantitative measurements or limits on jet quenching effects in such collisions are still lacking.  

In this paper we present measurements sensitive to jet quenching in \pPb\ collisions at 
$\sqrtsNN =5.02$\,TeV, based on the semi-inclusive distribution of charged
jets recoiling from a high-\pT\ trigger hadron~\cite{deFlorian:2009fw}. The 
observable used in this analysis has been measured in \pp\ collisions at 
$\sqrts=7$\,TeV and compared to calculations based on PYTHIA and on pQCD 
at NLO, with PYTHIA providing a 
better description~\cite{Adam:2015pbpb}. It has also been used to measure jet 
quenching effects in 
\PbPb\ collisions at $\sqrtsNN=2.76$\,TeV \cite{Adam:2015pbpb} and in \AuAu\ 
collisions at $\sqrtsNN=200$\,GeV \cite{Adamczyk:2017yhe}.

The semi-inclusive recoil jet distribution is equivalent to the ratio 
of inclusive cross sections \cite{Adam:2015pbpb}; comparison of such self-normalized 
coincidence distributions for \pPb\ event populations with different EA therefore does not require 
scaling by the nuclear overlap function $\left<\TpPb\right>$. Measurement of this 
observable in \pPb\ collisions is sensitive to jet quenching effects, and 
indeed does not require interpretation of the EA in terms of \pPb\ collision 
geometry. This approach thereby avoids potential bias due to Glauber 
modeling when interpreting the measurement.

We report charged recoil jet distributions reconstructed with the \antikT\ 
algorithm \cite{FastJetAntikt} in the range 
$15<\pTjetch<50$\,\gev, for jet resolution parameters $\rr=0.2$ and 0.4.
Correction of the jet yield for background uncorrelated with the triggered hard process, including multi-partonic interactions (MPI), is 
carried out statistically at the level of ensemble-averaged 
distributions, using the data-driven method first applied in 
\cite{Adam:2015pbpb}. EA in \pPb\ collisions is characterized by two different 
observables, forward charged-particle 
multiplicity and neutral energy along the beam axis, both measured in the 
direction of the Pb-beam \cite{Abelev:2015cnt}. 
Jet quenching effects are quantified by comparing the measured distributions in 
different EA classes of the \pPb\ dataset. The results are compared to other jet quenching 
measurements and to theoretical calculations of jet quenching in asymmetric 
collision systems.

The paper is organized as follows: Sect.~\ref{sect:DatasetOffline} describes the 
data set and analysis; Sect.~\ref{sect:EvtSelect} describes event 
selection based on trigger hadrons and event activity; Sect.~\ref{sect:Jets} 
describes jet reconstruction; Sect.~\ref{sect:Observable} discusses the 
semi-inclusive observable and presents the raw data; Sect.~\ref{sect:Corr} 
discusses corrections; Sect.~\ref{sect:Uncert} discusses systematic 
uncertainties; Sect.~\ref{sect:Results} presents results and discussion; 
Sect.~\ref{sect:Comparison} compares the results to other measurements; and 
Sect.~\ref{sect:Summary} is the summary.

\section{Data set and analysis}
\label{sect:DatasetOffline}

The ALICE detector and
performance are described in \cite{Aamodt:2008zz, Abelev:2014ffa}. The data used 
in this analysis were recorded during the 2013 LHC run with \pPb\ collisions 
at $\sqrtsNN=5.02$\,TeV. The Pb-going 
direction has rapidity $y>0$ and pseudorapidity $\eta>0$ in the laboratory frame.
The per-nucleon momenta of the 
beams in this run were imbalanced in the laboratory frame, with the 
nucleon-nucleon center-of-mass at rapidity 
$\yNN = -0.465$. The acceptance of tracks and 
jets in this analysis are specified in terms of 
$\ystar=\yLAB-\yNN$, where \yLAB\ denotes the 
rapidity measured in the laboratory frame. 

Events were selected online by an MB trigger, which requires the
coincidence of signals in the V0A and V0C forward scintillator arrays. 
The V0A array has acceptance $2.8<\eta<5.1$ and the V0C array has acceptance 
$-3.7<\eta<-1.7$, both 
covering the full azimuth. Offline event selection also utilizes the Zero-Degree 
Calorimeters (ZDC), which are neutron calorimeters at zero degrees relative to the beam 
direction, located at a distance 112.5\,m from the nominal interaction point. The ZDC in the Pb-going direction is labeled ZNA.

Jet reconstruction in this analysis uses charged-particle tracks. Tracks are measured by the Inner Tracking System (ITS), a 
six-layer silicon vertex tracker, and the Time Projection Chamber (TPC). The 
tracking system acceptance covers $|\eta| < 0.9$ over the full azimuth, with tracks reconstructed 
in the range $0.15<\pT<100$\,\gev. Primary vertices are reconstructed offline by
extrapolation of these tracks to the beam axis.
Primary tracks are defined as reconstructed 
tracks with Distance of Closest Approach to the primary vertex in
the transverse plane $DCA_{xy} < 2.4$\,cm. 

The analysis uses 
high-quality primary tracks that include at least one track point in the Silicon 
Pixel Detector (SPD), 
which comprises the two 
innermost layers of ITS. The azimuthal distribution of such high-quality tracks 
is non-uniform, however, due to the non-uniform acceptance of the SPD in this run. Azimuthal 
uniformity in the tracking acceptance is achieved by supplementing the 
high-quality tracks 
with complementary tracks that do not have a hit in the SPD, 
which constitute 4.3\% of all primary tracks. The momentum resolution of 
complementary tracks, without an 
additional constraint, is lower than that of high-quality tracks. Complementary 
tracks are 
therefore refit, including the reconstructed primary vertex as a track point. 
Tracking efficiency for primary tracks is about 81\% for $\pT>3$\,\gev. 
Primary-track momentum resolution is 0.7\% at $\pT=1$\,\gev, 1.6\% at 
$\pT=10$\,\gev, and 4\% at $\pT=50$\,\gev.
Further details on the track selection and tracking performance in this analysis 
are given in \cite{Abelev:2013kqa,Adam:2016ppb}.

The MB trigger efficiency for non-single diffractive (NSD) collisions is 
$97.8\pm 3.1$\% \cite{Abelev:2013ppbtrg, Adam:2016ppb}. Since this is a correlation 
analysis, no correction is applied for the trigger inefficiency. 
Timing cuts on the V0 and ZDC signals, which are applied offline,
remove background events with vertices outside of the 
nominal \pPb\ interaction region that arise from beam-gas interactions and 
interactions with satellite beam bunches~\cite{Abelev:2014ffa}. 

Event pileup, due to multiple interactions
in the triggered bunch crossing, is suppressed by rejecting events with multiple 
primary vertex candidates.
For this procedure, a new set of primary vertex candidates is 
constructed from tracklets constructed solely from SPD hits (``SPD vertices").
SPD vertices have at least five SPD tracklets within  $DCA<1$\,mm
 and lie within the expected envelope of \pPb\ interaction points, with a 
distance not more than $3\sigma$ in $z$ or $2\sigma$ in the $xy$ plane from the 
centroid of the distribution. The minimum distance in $z$ between SPD vertices 
is 8\,mm. Events with multiple SPD vertices are rejected from further analysis. 
	The EA-bias of the pileup rejection procedure is negligible, due to the large separation of pileup vertices in $z$ and the requirement that each SPD vertex have at least 
five contributors. In this dataset, the average number of
 interactions per bunch crossing was $\mu\approx 0.3$--0.5\%, and this
pileup rejection procedure removes less than 0.15\% of all events.
 
In addition, accepted events must have the primary vertex (defined above) with
$|z_{\rm{vtx}}|<10$\,cm relative to the nominal center of the ITS along the beam 
axis. After all event selection cuts, the number of events in 
the analysis is $96\times10^{6}$, corresponding to an integrated luminosity of 
46\,$\mu {\mathrm{b}}^{-1}$.

Simulations are used to correct the raw data for instrumental effects, and to 
compare the corrected measurements to expectations from an event generator.
Simulated events were generated for \pp\ collisions at $\sqrts = 5.02$\,TeV 
using PYTHIA 6.425 with the Perugia 11 tune \cite{Skands:2010ak}. These events, 
labeled ``particle-level," include all primary charged particles as defined in 
\cite{PublicNotePrimaryPart}. Following the procedure in \cite{Adam:2015hoa}, 
instrumental effects are calculated by passing particle-level events through a 
detailed model of the ALICE detector based on GEANT3 \cite{GEANT3}.
These events are reconstructed with the same procedures that are used for real 
data; the output of this process is labeled ``detector-level." 
Comparison to data also uses a particle-level simulation of \pp\ collisions at 
$\sqrts=5.02$\,TeV generated with PYTHIA 8.215 Tune 4C~\cite{Pythia8:2007}.
All simulations take account of the
nucleon-nucleon center-of-mass rapidity shift of the \pPb\ data.

\section{Event selection}
\label{sect:EvtSelect}

This analysis is based on the semi-inclusive distribution of 
jets recoiling from 
a high-\pT\ trigger hadron. Event selection requires the presence of a 
high-\pT\ charged track, called the Trigger Track (TT), in a specified \pTtrig\ 
interval. Two exclusive event sets are defined, based on different TT intervals: 
$12<\pTtrig<50$\,\gev, denoted TT\{12,50\}, and $6<\pTtrig<7$\,\gev, denoted 
TT\{6,7\}. 

The choice of the upper TT interval limits is driven by two competing factors: the hardening 
of the recoil jet \pT-spectrum with increasing \pTtrig, and the decrease of 
the inclusive hadron production cross section for increasing \pTtrig. The 
choice of TT\{12,50\} provides the optimum kinematic reach and 
statistical precision of the normalized recoil jet spectrum for this dataset. 
The criteria for the lower TT interval, TT\{6,7\}, are that it be significantly 
lower in \pTtrig, with correspondingly softer recoil jet spectrum, while still 
in the region in which inclusive hadron production can be well-described 
perturbatively using collinear fragmentation functions~\cite{d'Enterria:2013vba, deFlorian:2014xna}.

The fraction of such events in the MB population is $6.9\times10^{-4}$
for TT\{12,50\} and $6.4\times10^{-3}$ for TT\{6,7\}. However, an event may 
satisfy 
both the TT\{6,7\} and TT\{12,50\} selection criteria, since fragmentation of an 
energetic jet can generate hadrons in both TT selection intervals. A procedure 
is required to ensure exclusive, statistically independent 
datasets for the two TT-selected populations.
In addition, optimization of the statistical precision of the analysis requires 
similar number of events in the two TT classes. The MB population was 
therefore divided randomly into two subsets, whose sizes are inversely 
proportional to the 
relative rate of the two TT selections: 90\% of MB events are assigned to the 
TT\{12,50\} analysis, with the remaining 10\% assigned to the 
TT\{6,7\} analysis. 

An event can also contain multiple hadrons within a 
single TT interval, likewise arising from jet fragmentation. 
For events with at least one hadron satisfying TT\{6,7\}, the relative rate of 
two or more hadrons in an event satisfying TT\{6,7\} is 2.3\%; the corresponding 
relative rate of multiple hadrons satisfying TT\{12,50\} is 5.3\%. If an event 
contains more than one track in the assigned TT interval, 
the trigger hadron is chosen as the candidate with the highest \pT. The 
resulting \pT-distribution of trigger tracks is consistent with the shape of the inclusive hadron distribution within 2\%. After the TT event selection procedure there are 63k events accepted that satisfy TT\{6,7\} and 60k events accepted that satisfy 
TT\{12,50\}.

A different procedure was 
employed in~\cite{Adam:2015pbpb} for the case of multiple trigger candidates in 
a TT interval, 
where the trigger track was chosen randomly 
amongst the candidates. However, the analysis reported here has a wider range in 
\pT\ for the upper TT class, and random selection results in reduced level of 
agreement ($\sim$10\%) of the trigger track \pT-distribution with the inclusive 
hadron spectrum shape. The full analysis was also 
carried out for this choice of procedure for trigger selection, and all 
resulting physics distributions agree with those of the primary analysis within 
the
uncertainties.

Measurement of EA uses signals from V0A and ZNA.
Classification of events in percentile intervals of the V0A and ZNA 
signal distributions is discussed in \cite{Abelev:2015cnt}. 

About 5\% of accepted events do not have a ZNA signal above the detector 
threshold. The ZNA threshold is set so that the detector is fully efficient for 
single neutrons. However, these 
events correspond to \pPb\ collisions in which the Pb-nucleus remnant is not 
accompanied by any beam-rapidity single neutrons. The 
distribution of mid-rapidity track multiplicity for these events resembles 
closely that for events with low but observable ZNA signal, and these events are 
therefore assigned to the bin with lowest ZNA signal.

\begin{figure}[tbh!]
\centering
\includegraphics[width=0.49\textwidth]{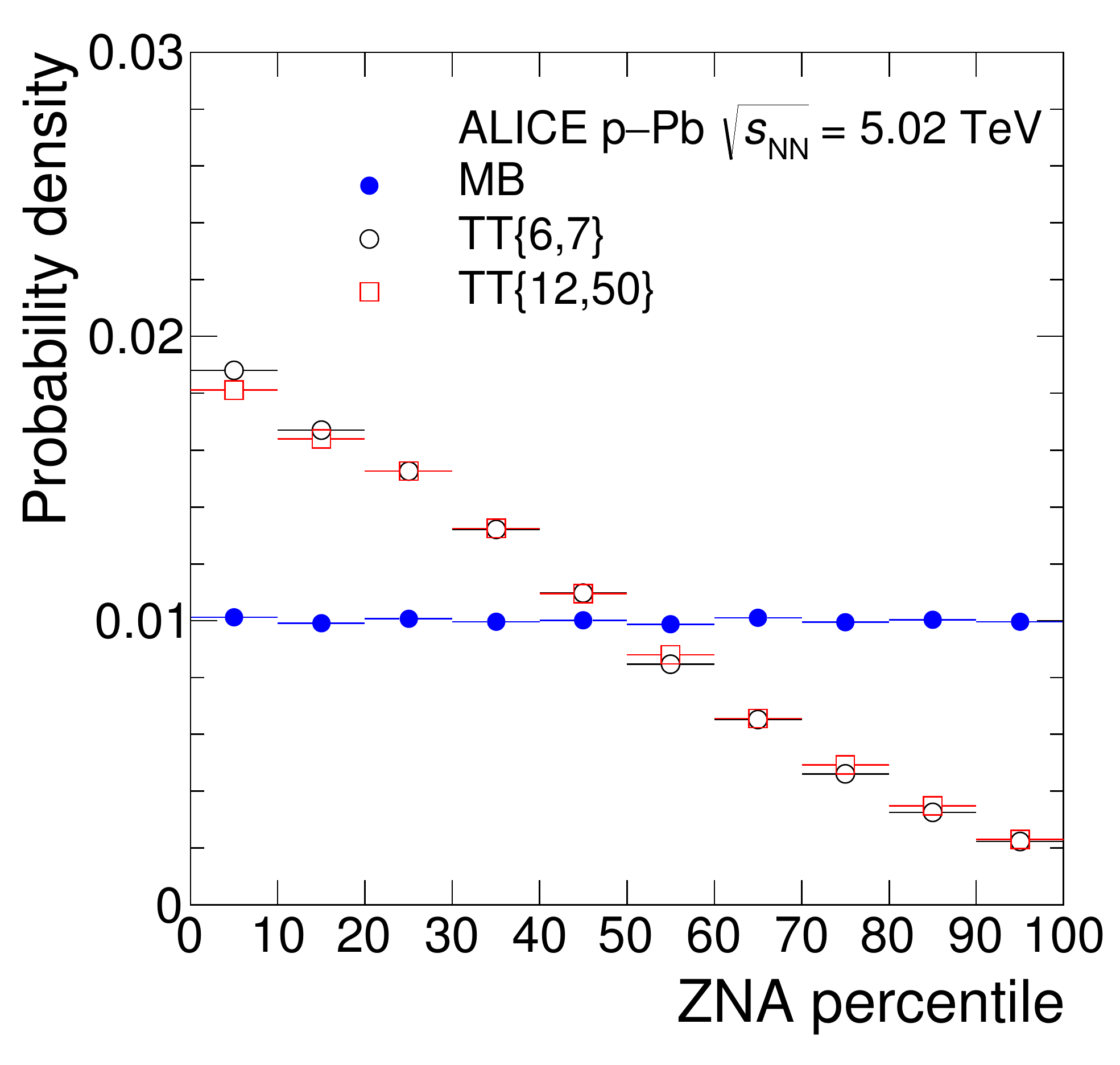}
\includegraphics[width=0.49\textwidth]{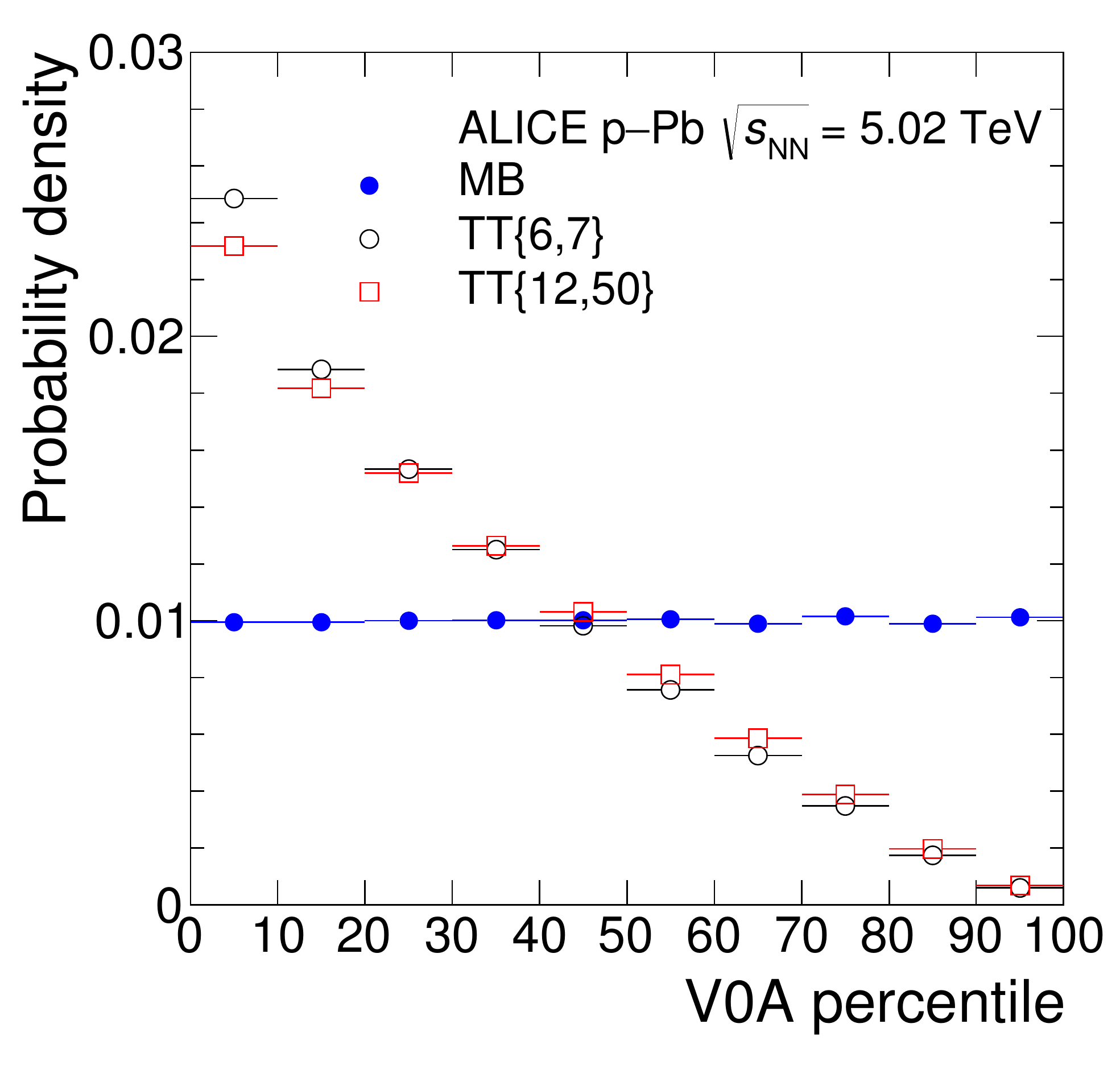}
\caption{Distribution of event activity EA in decile bins measured 
in ZNA 
(left) and V0A (right), for the MB event population and for event populations 
selected with the requirement of a high-\pT\ trigger hadron in the intervals 
$6<\pTtrig<7$\,\gev\ (TT\{6,7\}) and $12<\pTtrig<50$\,\gev\ (TT\{12,50\}). 
Large EA is to the left, with the 0-10\% bin representing 
the largest EA, or highest amplitude signal in ZNA or V0A.}
\label{fig:EAbias}
\end{figure}

Figure~\ref{fig:EAbias} shows the distribution of EA measured by ZNA and V0A, 
in decile bins of signal amplitude. The 
decile bin limits are determined from their distributions in the 
MB population, with MB events therefore distributed uniformly in this projection 
by construction. The figure also shows V0A and ZNA distributions for event 
populations selected by the TT\{6,7\} and the TT\{12,50\} criteria. Requiring the presence of a high-\pT\ hadron trigger in an event is seen to induce a 
bias 
towards larger EA, corresponding to larger amplitude in both ZNA and V0A. A 
small dependence on the TT class (i.e. on \pTtrig) is also observed, with 
magnitude less than 10\% of the overall bias, and with the TT-dependence 
slightly larger for V0A than for ZNA. Figure~\ref{fig:EAbias} shows 
significant correlation between EA and the presence of a hard 
process in the central region.

For further analysis, events were assigned to wider percentile bins in ZNA or 
V0A, based on their MB distributions: 20\% of the MB 
population with largest signal (``0--20\%"), the next 30\% 
(``20--50\%"), and 
the remaining 50\% with the lowest signal (``50--100\%").
The bias imposed by TT selection, shown in 
Fig.~\ref{fig:EAbias}, corresponds to different fractions of the 
TT-biased population: the nominal 0--20\% ZNA interval corresponds to 0--35\% 
of the 
TT-biased population; the nominal 20--50\% ZNA interval corresponds to 35--74\% 
of TT-biased; and the nominal 50--100\% ZNA interval corresponds to 74--100\% 
of TT-biased. Similar modification of percentile fractions due to TT bias is observed 
for the V0A signal. 

The same events are used for the ZNA and V0A selections, so 
that the analyses using the two different 
EA metrics are not statistically independent. 

\section{Jet reconstruction}
\label{sect:Jets}

Several types of jet are used in the analysis, which we distinguish by the
notation for jet \pT: \pTraw\ refers to the output of the jet reconstruction 
algorithm; \pTreco\ is \pTraw\ after subtraction of an estimated contribution to jet \pT\
of uncorrelated background; and 
\pTjetch\ refers to the fully corrected jet spectrum. For simulations, \pTgen\ 
refers to reconstructed charged-particle jets at the particle-level, and \pTdet\ 
refers to reconstructed charged-particle jets at the detector-level.

Jet reconstruction is carried out using the \kT\ and \antikT\ algorithms 
\cite{FastJetAntikt} with the boost-invariant \pT\ recombination scheme \cite{Cacciari:2011ma}, using all accepted charged tracks with 
$\pT>0.15$\,\gev. Jet area \Ajet\ 
is calculated using the Fastjet algorithm 
\cite{FastJetArea} with ghost area 0.005. 

Two jet reconstruction passes are carried out for each event. The first pass 
estimates the level of uncorrelated background energy in the event, while the second pass 
generates the set of jet candidates used in the physics analysis, with 
adjustment of their \pT\ using the estimated background level from the first pass. 

In the first pass, the \pTraw\ distribution reconstructed  by the \kT\ algorithm with $\rr =0.4$ is used to estimate $\rho$, the magnitude of background energy per unit area~\cite{FastJetPileup},

\begin{equation}
\rho = {\mathrm{median}}_{\kT\ {\mathrm{jets}}}\left\{ \frac{\pTraw\ }{ \Ajet\ 
}\right\}, 
\label{eq:rho}
\end{equation} 

\noindent
where the median is calculated by excluding the jet which has the trigger hadron as a constituent. A different $\rho$ estimator~\cite{Chatrchyan:2012bg} is utilized to assess the systematic uncertainties of this procedure.

The second jet reconstruction pass is carried out using the \antikT\ algorithm with $\rr = 0.2$ and 0.4. The value of \pTraw\ for each jet candidate from this step is then adjusted for 
the estimated background energy density~\cite{FastJetPileup},

\begin{equation}
\pTreco =\pTraw - \Ajet\cdot\rho.
\label{eq:pTreco}
\end{equation}

\noindent
A jet candidate from the second pass is accepted for further analysis if its area satisfies $\Ajet > 0.6 \pi \rr^{2}$~\cite{Abelev:2013kqa,Adam:2015pbpb}, and its jet axis lies within $|\eta_{\rm jet}|<0.9-\rr$ and an azimuthal interval situated back-to-back with respect to the TT, $\dphi_{\mathrm{recoil}}>\pi-0.6$, where $\dphi = \varphi_{\rm TT} - \varphi_{\rm jet}$ and $0<\dphi<\pi$. An event may have multiple accepted jet candidates. 

For further analysis we follow the procedure used in~\cite{Adam:2015pbpb}, in 
which no 
rejection of individual jet candidates is carried out. Recoil jet distributions 
are accumulated for the selected event 
populations, and corrections for uncorrelated jet yield and for smearing and 
residual shift of 
\pTjetch\ due to uncorrelated background are carried out at the level of 
the ensemble-averaged distributions, as discussed below.

Jet energy resolution due to instrumental effects (JER) and jet energy scale 
(JES) uncertainty are similar to those in~\cite{Adam:2015pbpb}. The JER is 
determined by comparing simulated jets at the particle and detector levels. The 
distribution of ($\pTdet-\pTgen $)/\pTgen\ is asymmetric, with a sharp peak 
centered at zero and a tail to negative values~\cite{Adamczyk:2017yhe}. Fit of a 
Gaussian function to the sharp peak gives $\sigma\simeq2$--3\%, while the full 
distribution has ${\rm RMS}\simeq25$\%, with both quantities having no significant 
dependence on \pTpart\ and \rr. The JES uncertainty, which is due predominantly 
to uncertainty in tracking efficiency, is 4\%, likewise with no significant 
dependence on \pTjetch\ and \rr. However, these values of JER and JES 
uncertainty, while helpful to characterize the jet measurement, are not used 
in the analysis. Corrections are carried out utilizing the full 
response matrix, which incorporates detailed distributions of all contributions 
to JER and JES uncertainty. The systematic uncertainties 
(Tab.~\ref{tab:SysUncert}) likewise take such factors fully into account.

\section{Observable and raw data}
\label{sect:Observable}
   
The semi-inclusive h+jet distribution corresponds to the \pT-differential distribution of recoil jets normalized by the number of trigger hadrons, \NtrigX, 
     
\begin{equation}
    \frac{1}{\NtrigX\ }\frac{{\mathrm d}^{2}\Njets\ }{ {\mathrm d}\pTjetch\ 
{\mathrm d}\etajet\ }\bigg|_{\substack{\pTtrig\ \in {\mathrm{TT}} \\
\dphi\in\mathrm{recoil}}} 
=\frac{1}{\sigma^{\mathrm{pPb}\rightarrow \mathrm{h+X}}}\frac{{\mathrm 
d}^{2}\sigma^{\mathrm{pPb}\rightarrow{\mathrm{h+jet+X}}}}{ {\mathrm d}\pTjetch\ 
{\mathrm d}\etajet\ }\bigg|_{\substack{\mathrm{h} \in {\mathrm{TT}} \\ \dphi\in\mathrm{recoil}}} 
  \label{eq:RecoilJetYield}
\end{equation}

\noindent
All accepted jets contribute to the distribution on the LHS. This distribution is 
equivalent to measurement of the ratio of two cross sections, as shown on the RHS: the 
coincidence cross section 
for both trigger hadron and recoil jet to be in the acceptance, divided by the 
inclusive production cross section for trigger hadrons. This expression applies 
to both the MB event population, and to event subsets selected by EA. The 
features of this observable and its theoretical calculations are discussed in 
detail in Refs.~\cite{Adam:2015pbpb,Adamczyk:2017yhe}. Here we consider two 
specific aspects of this distribution.

The first aspect is the bias imposed by the high-\pT\ hadron trigger. 
For collision systems in which jet quenching occurs, selection of high-\pT\ 
hadrons is thought to bias towards the fragments of jet that have experienced 
little quenching, due to the combined effect of jet energy loss and the shapes 
of the inclusive jet production and the jet fragmentation 
distributions~\cite{Baier:2002tc,Drees:2003zh,Dainese:2004te,Eskola:2004cr,Renk:2006nd,Loizides:2006cs,Zhang:2007ja,Renk:2012cb}. 
If that is the case, then the hadron trigger 
bias in this measurement would be independent of EA. 
This conjecture is supported by ALICE measurements of inclusive hadron 
production in \pPb\ 
collisions that find no significant yield
modification in the trigger \pT-range of this 
measurement, for both the MB and EA-selected event 
populations~\cite{Abel:2015ppb,Abelev:2015cnt}. 
The picture provided by current ATLAS and CMS hadron production 
measurements~\cite{Khachatryan:2015xaa,Aad:2016zif} is more 
complex, however. Further study of this conjecture
requires additional measurements of inclusive hadron production in \pp\ and 
\pPb\ collisions, together with theoretical calculations incorporating jet 
quenching that accurately reproduce these measurements.

The second aspect is the effect of trigger hadron efficiency on the 
equality in Eq.~\ref{eq:RecoilJetYield}. As noted in Sect.~\ref{sect:EvtSelect}, 
the analysis requires selection of a single trigger hadron in each event. 
However, in a few percent of events there are multiple hadrons satisfying the TT 
selection criteria, of which only one is chosen as trigger. Consequently, not 
all hadrons that would contribute to measurement of the inclusive hadron cross 
section (first term on the RHS of Eq.~\ref{eq:RecoilJetYield}) also contribute 
to \NtrigX\ (first term on the LHS of Eq.~\ref{eq:RecoilJetYield}). However, as 
noted above, the shape of the trigger hadron \pT-distribution is consistent 
with that of the inclusive hadron spectrum within 2\%. In other words, the 
trigger distribution used in practice samples the inclusive hadron distribution 
with efficiency less than unity but without \pT-dependent bias, within a 
precision of 2\%. This same inefficiency also applies to the h+jet coincidence 
process in the second term on the LHS of Eq.~\ref{eq:RecoilJetYield}, and it 
therefore cancels identically in the ratio. Equation~\ref{eq:RecoilJetYield} 
therefore remains 
valid for trigger selection efficiency less than unity.

The study of jet quenching using inclusive yields requires 
comparison of the inclusive distribution measured in heavy ion collisions
to a reference distribution measured in a system in which quenching effects 
are not expected, usually \pp\ collisions at the same 
\sqrtsNN. Such comparisons must account for the effect of multiple 
nucleon-nucleon collisions in each collision of heavy nuclei, which arises 
due to nuclear geometry. For inclusive distributions in \pPb\ 
collisions this is 
accomplished by scaling inclusive cross 
sections for \pp\ collisions by $\left<\TpPb\right>$, 
which is calculated by modeling based on Glauber theory 
under the 
assumption that EA is correlated with the collision 
geometry~\cite{Miller:2007ri,Adams:2003im,Adler:2006wg, 
Abel:2015ppb,Abelev:2015cnt, 
Aad:2016zif,Khachatryan:2015xaa,Adam:2015hoa,Adam:2016ppb, 
Adare:2016dau,Aad:2015ppb}. 

For the semi-inclusive distribution in Eq.~\ref{eq:RecoilJetYield}, the 
reference distribution without nuclear effects is

\begin{align}
    \frac{1}{\sigma_\mathrm{ref}^{\mathrm{pPb}\rightarrow 
\mathrm{h+X}}}\frac{{\mathrm 
d}^{2}\sigma_\mathrm{ref}^{\mathrm{pPb}\rightarrow{\mathrm{h+jet+X}}}}{ {\mathrm 
d}\pTjetch\ {\mathrm d}\etajet\ }\bigg|_{\substack{h \in {\mathrm{TT}} \\ \dphi\in\mathrm{recoil}}} 
    &= 
\frac{1}{\left<T_{\mathrm{pPb}}\right>\cdot\sigma^{\mathrm{pp}\rightarrow\mathrm{h+X}}}\frac{\left<T_{\mathrm{pPb}}\right>\cdot{\mathrm 
d}^{2}\sigma^{\mathrm{pp}\rightarrow\mathrm{h+jet+X}}}{ {\mathrm d}\pTjetch\ 
{\mathrm d}\etajet\ }\bigg|_{\substack{h \in {\mathrm{TT}} \\ \dphi\in\mathrm{recoil}}}  
\nonumber \\  
    &= \frac{1}{\sigma^{\mathrm{pp}\rightarrow\mathrm{h+X}}}\frac{{\mathrm 
d}^{2}\sigma^{\mathrm{pp}\rightarrow\mathrm{h+jet+X}}}{ {\mathrm d}\pTjetch\ 
{\mathrm d}\etajet\ }\bigg|_{\substack{\mathrm{h} \in {\mathrm{TT}} \\ \dphi\in\mathrm{recoil}}}        
  \label{eq:sigma}
\end{align} 

\noindent
Since the scaling factors $\left<\TpPb\right>$ in the numerator and denominator cancel 
identically, 
the reference distribution for this observable has no dependence 
on $\left<\TpPb\right>$. In other words, this distribution is self-normalized, and measurement of jet quenching using this observable does not 
require 
Glauber modeling for the reference spectrum. In particular, the assumption that 
event activity is correlated with the collision 
geometry is not required. 

A similar approach, utilizing a coincidence observable to measure jet quenching 
in high-mulitplicity \pp\ collisions, was recently proposed 
in~\cite{Mangano:2017plv}.
   
\begin{figure}[tbh!]
\centering
\includegraphics[width=0.575\textwidth]{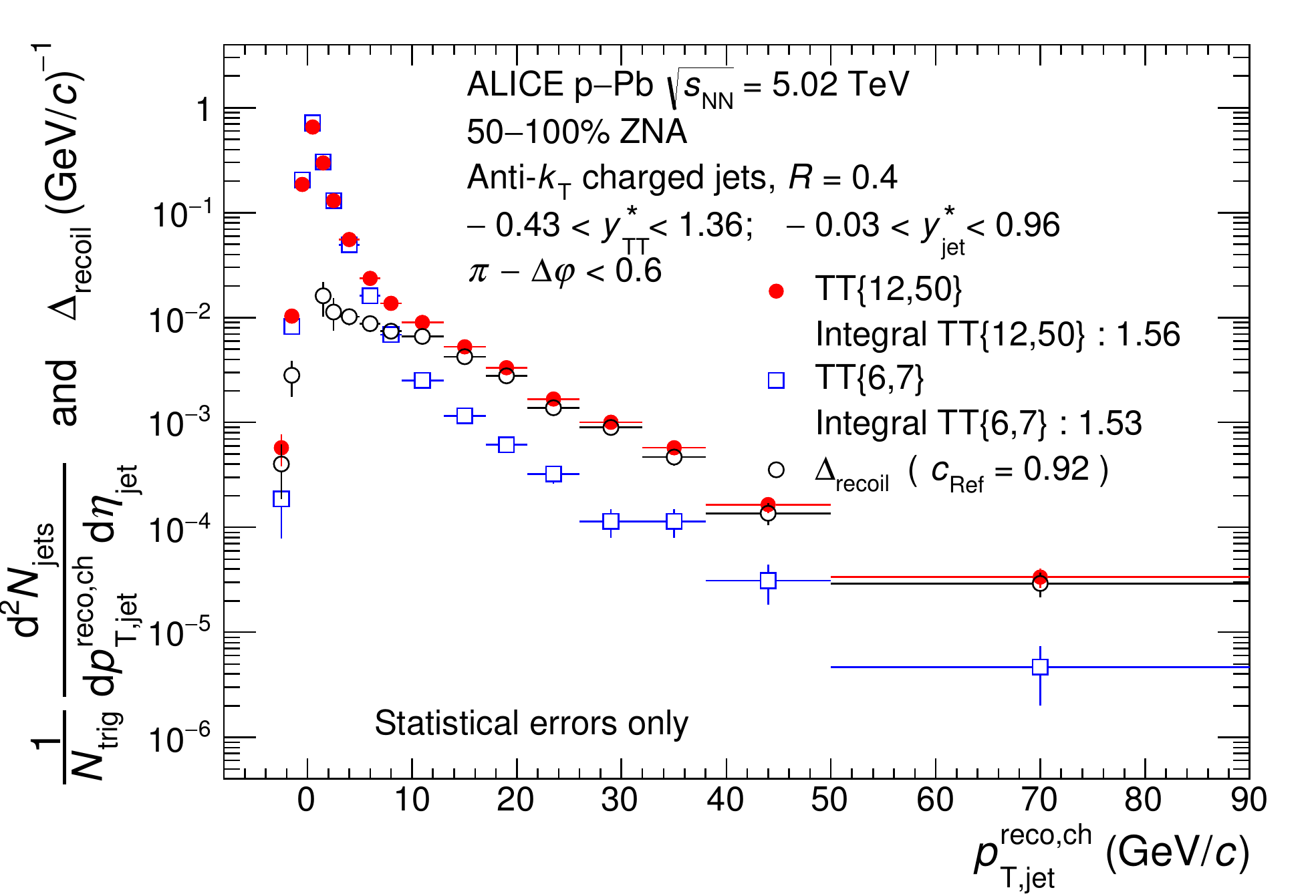}
\includegraphics[width=0.41\textwidth]{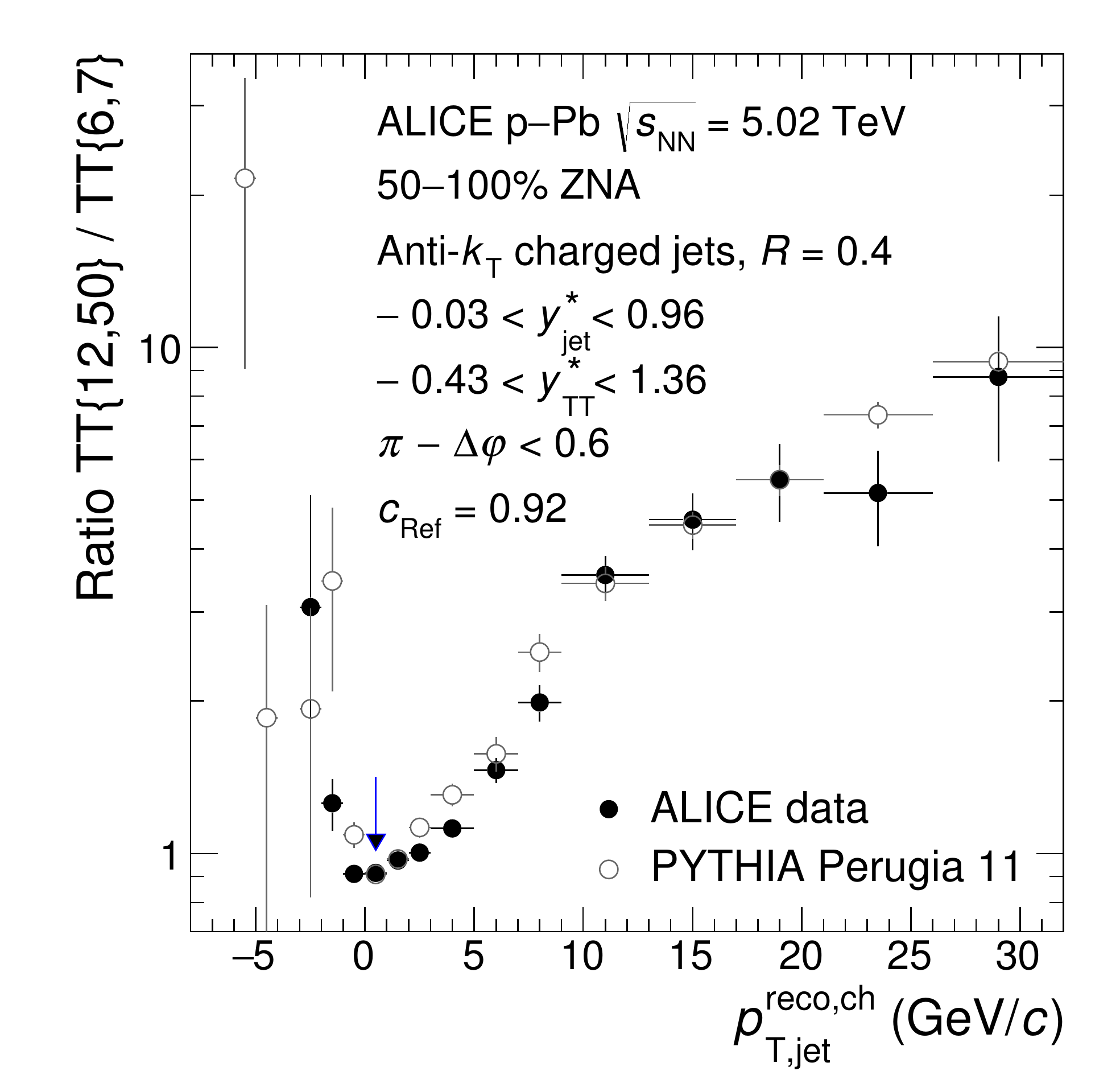}
\includegraphics[width=0.575\textwidth]{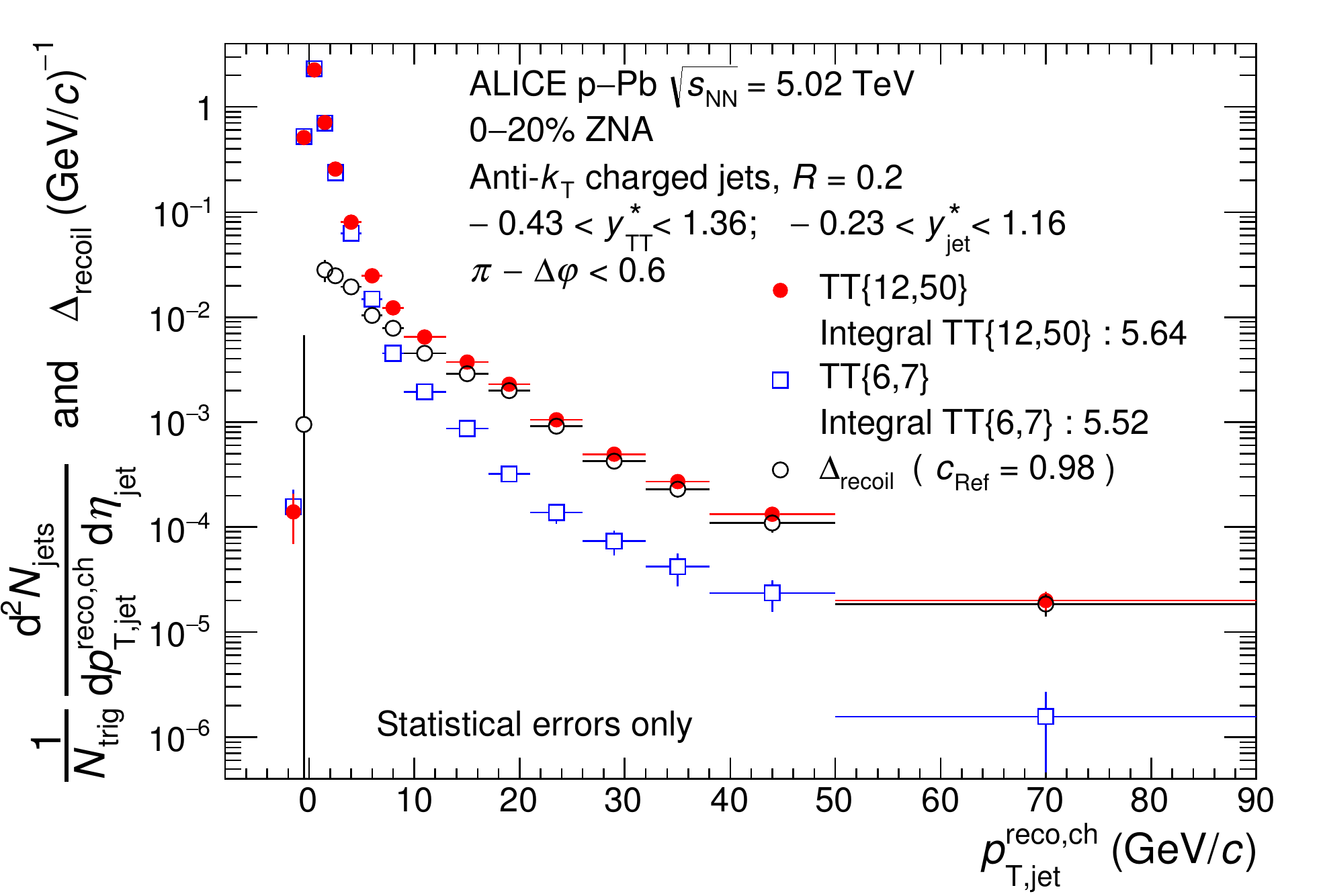}
\includegraphics[width=0.41\textwidth]{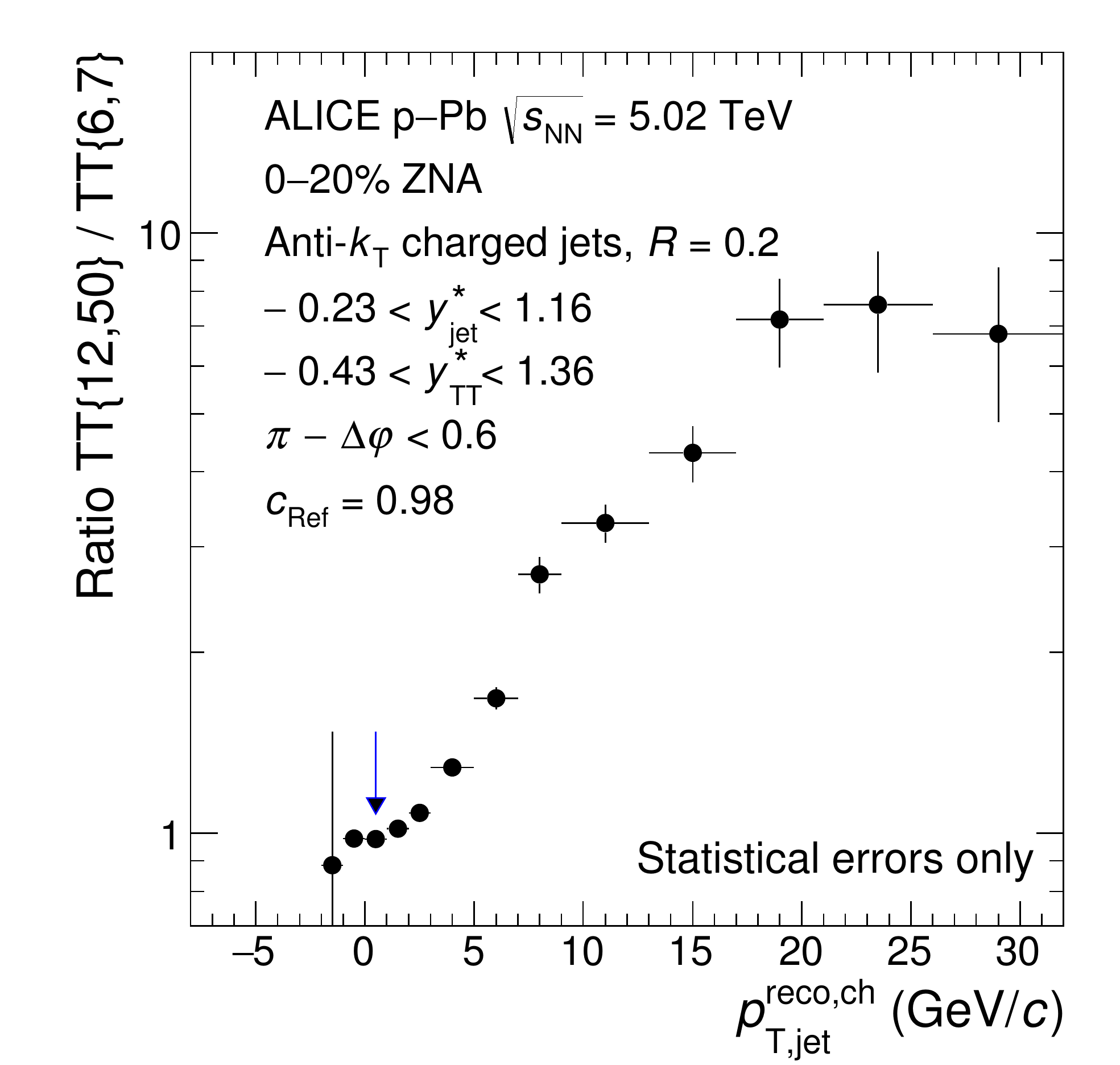}
\includegraphics[width=0.575\textwidth]{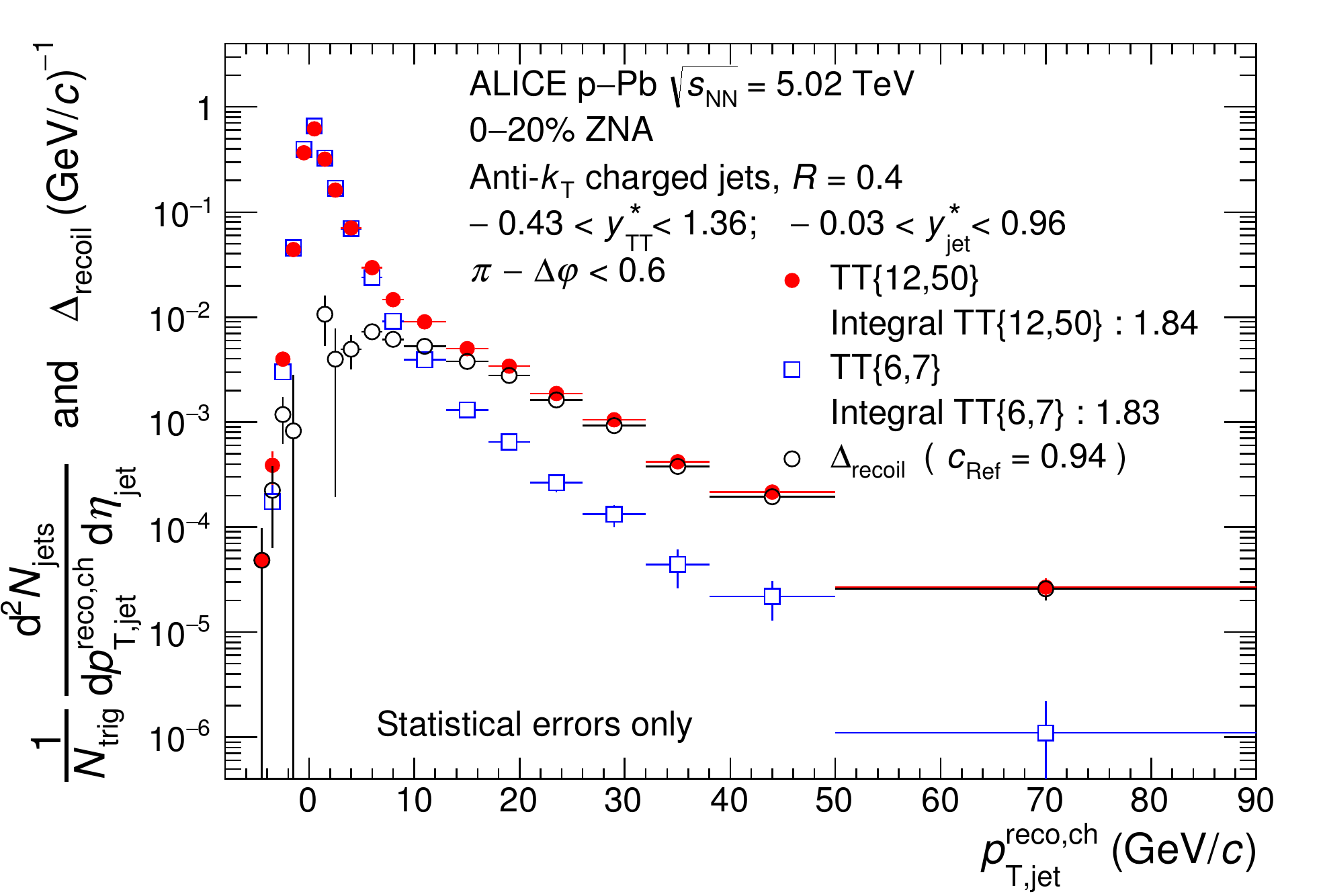}
\includegraphics[width=0.41\textwidth]{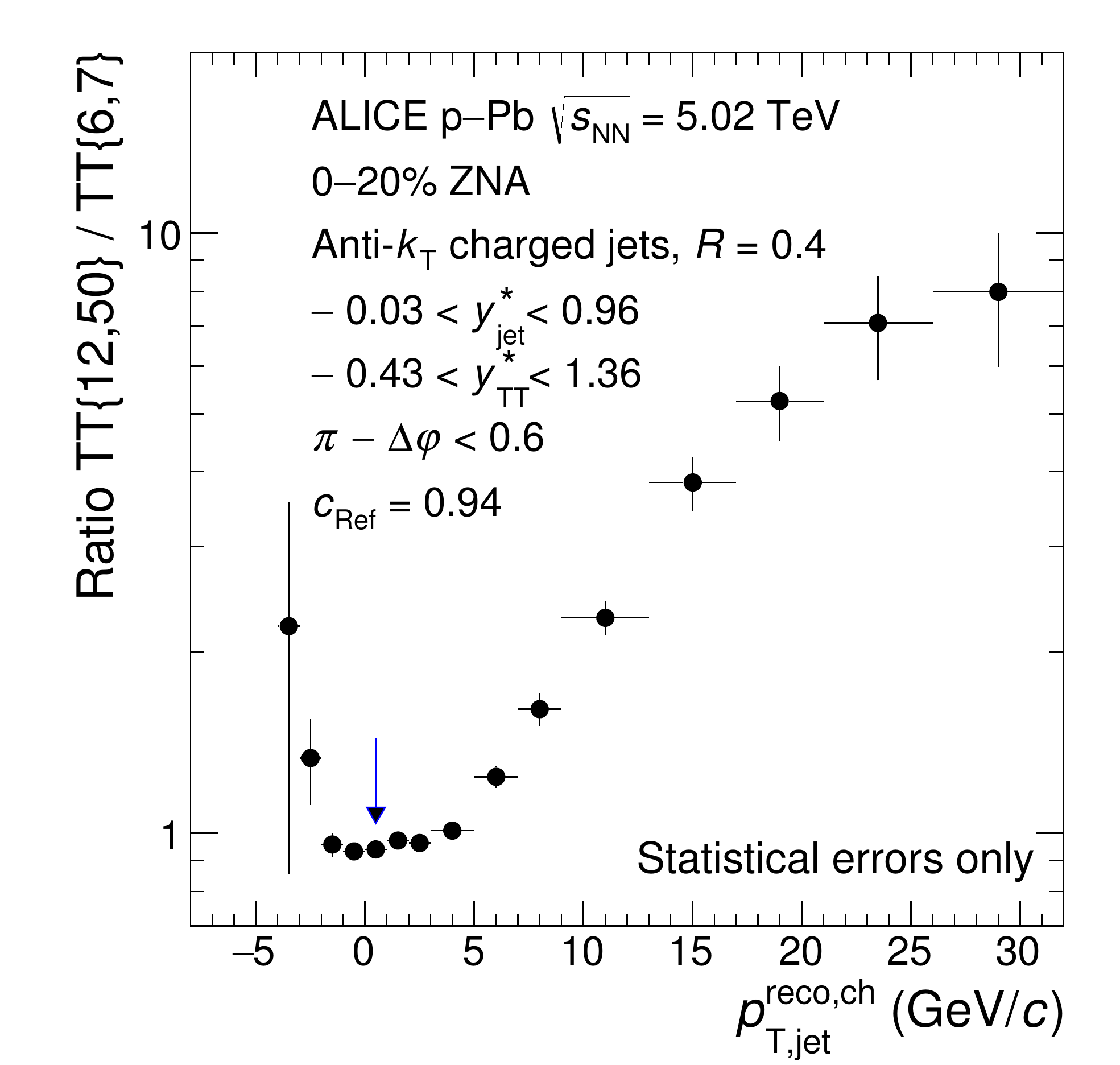}
\caption{Uncorrected semi-inclusive distributions of 
  charged jets recoiling from a high-\pT\ hadron
  trigger in \pPb\ collisions at $\sqrtsNN = 5.02$\,TeV with the EA selection of 50--100\% in ZNA for $\rr = 0.4$ (top panels), and with the EA selection of 0-20\% in ZNA for $ \rr = 0.2$ (middle panels) and $\rr = 0.4$ (bottom panels). The acceptance for TT and recoil jets in the CM frame are denoted \ystarTT\ and \ystarjet, respectively. Left panels: raw distributions for TT\{12,50\} (red circles) and TT\{6,7\} (blue boxes), and the corresponding \Drecoil\ distribution (Eq.~\ref{eq:Drecoil}, black circles). Right panels: ratio of yields for TT\{12,50\}/TT\{6,7\} measured by ALICE in \pPb\ collisions and calculated using detector level PYTHIA Perugia 11 simulation of \pp\ collisions at $\sqrts = 5.02$\,TeV. The PYTHIA-generated ratios in the top right and bottom right panels are the same. The arrow indicates the 0--1\,\gev\ bin which is used to calculate \cRef. The uncertainties are statistical only.
}
\label{fig:hJetRecoilRaw}
\end{figure}

Figure~\ref{fig:hJetRecoilRaw}, left panels, show recoil-jet distributions for $\rr =0.4$ in \pPb\ collisions with the
50--100\% ZNA selection, and for $\rr =0.2$ and 0.4 with the 0--20\% ZNA selection. 
Distributions in EA intervals selected with V0A and with 20--50\% 
ZNA are similar~\cite{PublicNote}.
The distributions have non-zero  yield for 
$\pTreco<0$, because regions of an event can have energy density less than 
$\rho$~\cite{Adam:2015pbpb}. These distributions
are significantly narrower in the region $\pTreco<0$ than those observed 
in central \PbPb\ collisions, where the uncorrelated component of the event is 
significantly larger~\cite{Adam:2015pbpb}. 

Figure~\ref{fig:hJetRecoilRaw}, right panels, show ratios of the
distributions for the two TT classes. The right panels also 
show the corresponding ratio for
\pp\ collisions at $\sqrts = 5.02$\,TeV, using simulated detector-level events 
generated with PYTHIA Perugia 11. For $\pTreco\sim0$ the 
two distributions agree within $\sim10$\% for both values of \rr, consistent 
with the expectation that yield in this region 
arises predominantly from processes that are uncorrelated with the 
trigger hadron~\cite{Adam:2015pbpb}. At larger \pTreco, the distribution for 
TT\{12,50\} exceeds that for TT\{6,7\}. This dependence of the recoil 
distribution on \pTtrig\ is expected from QCD-based considerations, since 
higher \pTtrig\ biases towards hard processes with higher 
$Q^2$ on average. Indeed, hardening of the semi-inclusive recoil jet 
distribution with 
increasing \pTtrig\ is also seen in the PYTHIA-generated ratios for \pp\ collisions at $\sqrts = 5.02$\,TeV shown in the figure, 
and has been measured in \pp\ collisions at $\sqrts 
=7$\,TeV and observed in theoretical calculations based on NLO pQCD and on 
PYTHIA~\cite{Adam:2015pbpb}.

The PYTHIA-generated ratio for \pp\ collisions reproduces well the 
ratio measured for low-EA  \pPb\ collisions (ZNA 50--100\%, 
Fig.~\ref{fig:hJetRecoilRaw} upper right panel), while the level of agreement 
between the simulation and measurements is not as good for high-EA \pPb\ 
collisions (ZNA 0--20\%, Fig.~\ref{fig:hJetRecoilRaw}, middle and bottom right 
panels). This occurs because
there is larger uncorrelated background in high-EA than in low-EA \pPb\ 
collisions. 

The distribution of jet candidates that are uncorrelated with the trigger is independent of \pTtrig, by definition. The distribution of correlated recoil jets can 
therefore be measured using the \Drecoil\ observable, which is the difference of 
the two normalized recoil distributions \cite{Adam:2015pbpb},
     
\begin{equation}
  \Drecoil\left(\pTjetch\right) =  \frac{1}{\NtrigX\ }\frac{{\mathrm d}^{2}\Njets\ }{ {\mathrm 
d}\pTjetch\ }\bigg|_{\pTtrig\ \in \TTSig\ }
  - \cRef\ \cdot \frac{1}{\NtrigX\ }\frac{{\mathrm d}^{2}\Njets\ }{ {\mathrm 
d}\pTjetch\ }\bigg|_{\pTtrig\ \in \TTRef\ },
  \label{eq:Drecoil}
\end{equation}   

\noindent
where \TTSig\ and \TTRef\ refer to Signal and Reference TT intervals, in this 
analysis corresponding to TT\{12,50\} and TT\{6,7\} respectively. \Drecoil\ is 
normalized per unit \etajet, notation not shown. 

The Reference spectrum in \Drecoil\ is scaled by the factor \cRef\ to account 
for the invariance of the jet density with TT-class, as indicated by comparison of 
the spectrum integrals in Fig.~\ref{fig:hJetRecoilRaw}
and the larger yield of Signal spectrum at high 
\pTreco\ \cite{Adam:2015pbpb}. The value of \cRef\ in this analysis is taken as the ratio 
of the 
Signal and Reference spectra in the bin $0<\pTreco<1$\,\gev, as shown by the 
arrow in Fig.~\ref{fig:hJetRecoilRaw}, right panels. The value of 
\cRef\ lies 
between 0.92 and 0.99 for the various spectra. Additional variation in the value 
of \cRef\ was used to assess systematic uncertainties.

We note that the \TTRef\ distribution includes correlated recoil jet yield, so 
that the subtraction in 
Eq.~\ref{eq:Drecoil} removes both the trigger-uncorrelated yield and the 
\TTRef-correlated yield. The \Drecoil\ observable is therefore a differential, 
not absolute, measurement 
of the recoil spectrum~\cite{Adam:2015pbpb}, though the \TTRef\ component is 
significantly smaller than that in the \TTSig\ component over most of the 
\pTreco\ range.
The \Drecoil\ distributions in Fig.~\ref{fig:hJetRecoilRaw} lie significantly 
below the TT-specific distributions for $\pTreco<5$ \gev\ but 
agree with the TT\{12,50\} distribution within 15\% for $\pTreco>15$ \gev. 
These features indicate that the region of 
negative and small positive \pTreco\ is dominated by uncorrelated jet yield, 
while the region for large positive \pTreco\ is dominated by recoil jet yield 
that is correlated with \TTSig.

One contribution to uncorrelated background is jet yield due to Multiple 
Partonic Interactions (MPI), which can occur when two independent high-$Q^2$ 
interactions in the same \pPb\ collision generate the trigger hadron and a jet 
in the recoil acceptance. Since the two interactions are independent, the recoil 
jet distribution generated by MPI will be independent of \pTtrig, by definition, and will be removed from \Drecoil\ by the subtraction. No correction of \Drecoil\ for the 
contribution of MPI is therefore needed in the analysis.

The raw \Drecoil\ distributions, such as those in Fig.~\ref{fig:hJetRecoilRaw}, 
must still be corrected for jet momentum smearing due to instrumental effects and local background 
fluctuations, and for jet reconstruction efficiency. Jet quenching effects are 
measured by comparing the corrected \Drecoil\ distributions for different EA 
classes, and at different \rr.

\section{Corrections}
\label{sect:Corr}

Corrections for instrumental effects and local background fluctuations are carried out using unfolding methods \cite{Cowan:2002in,Hocker:1995kb,D'Agostini:1994zf}. The measured distribution \DrecoilM\ is related to the true distribution \DrecoilT\ by a linear transformation,

\begin{equation}
\DrecoilM(\pTreco) = \Rfull(\pTreco,\pTgen) \otimes \left[\mathrm{eff}(\pTgen)\cdot\DrecoilT(\pTgen)\right], 
\label{eq:Unfold}
\end{equation}

\noindent
where $\mathrm{eff}(\pTgen)$ is the jet reconstruction efficiency and \Rfull\ is 
the cumulative response matrix excluding jet reconstruction efficiency. The explicit specification of jet reconstruction efficiency in this expression, distinct from the unfolding step, makes interpretation of the unfolding procedure more transparent. \Rfull(\pTreco,\pTgen) is further assumed to factorize
as the product of separate response matrices 
for background fluctuations and instrumental response,

\begin{equation}
\Rfull(\pTreco,\pTgen) = \Rbkgd(\pTreco,\pTdet) \otimes \Rinstr(\pTdet,\pTgen).
\label{eq:RespMatrix}
\end{equation}

The matrix \Rfull\ can be close to singular, in which case the solution of 
Eq.~\ref{eq:Unfold} via direct inversion of \Rfull\ 
generates large fluctuations in central values and large variance due to the 
statistical variation in \DrecoilM(\pTreco) and \Rfull\ \cite{Cowan:2002in}. 
An approximate solution of Eq.~\ref{eq:Unfold} that 
is physically more meaningful is obtained by 
regularized unfolding, which imposes a smoothness 
constraint on the solution. Unfolding in this analysis is carried out using 
approaches based on 
Singular Value 
Decomposition (SVD)  \cite{Hocker:1995kb} and on Bayes' Theorem 
\cite{D'Agostini:1994zf}, as implemented in the 
RooUnfold package \cite{RooUnfold}. 

The instrumental response matrix, \Rinstr, is calculated from the simulated 
detector response applied to events generated by PYTHIA for \pp\ collisions at 
$\sqrts = 5.02$\,TeV. Jets at the particle-level and detector-level are 
matched in $(\eta,\phi)$ space by selecting the detector-level jet that is 
closest to the particle-level jet, and vice versa. An entry in \Rinstr\ is 
made for every matched pair. The \Rinstr\ matrix is normalized such 
that, for each bin in \pTpart, the sum over all bins in \pTdet\ is unity. 
In practice, however, the matching probability is less than unity, which is 
accounted for in Eq.~\ref{eq:Unfold} by the efficiency factor 
$\mathrm{eff}(\pTgen)$. No dependence of \Rinstr\ on EA of the \pPb\ 
event population was observed.

The background response matrix, \Rbkgd, is calculated by embedding single tracks 
with transverse momentum \pTembed\ into real \pPb\ events that contain a 
TT~\cite{Adam:2015pbpb}. The
relative azimuthal angle between the embedded track and the TT is in the range 
$\left[\pi/4,\,3\pi/4\right]$, to minimize overlap of the embedded track 
with the jet containing TT and with true recoil jets. These hybrid events are 
analyzed with the same procedures used for real data, and 
the jet containing the embedded track is identified. Smearing of jet candidate 
\pT\ due to background fluctuations is quantified by the distribution of

\begin{equation}
\dpT\ = \pTreco - \pTembed,
\label{eq:dpT}
\end{equation}

\noindent
where \pTreco\ refers to the jet containing the embedded track. \Rbkgd,  
the probability distribution of \dpT, is calculated separately for the MB 
population and for the various event populations selected by EA. Embedding of 
PYTHIA-generated jets rather than single tracks yields very similar \dpT\ 
distributions.

Unfolding follows the procedure described in \cite{Abelev:2013kqa}. The input to 
unfolding is the measured distribution \DrecoilM(\pTreco) in the 
range $1<\pTreco<90$\,\gev. The unfolding procedure requires specification of a prior distribution. For the 
primary analysis, the prior is the \Drecoil\ distribution calculated with 
PYTHIA8 tune 4C \cite{Pythia8:2007} for \pp\ collisions at 
$\sqrts =5.02$\,TeV. Regularization of SVD unfolding utilizes a statistical test to determine the 
transition between random fluctuations and statistically 
significant components of the $d$-vector \cite{Hocker:1995kb}, which is 
achieved typically with regularization parameter $k=4$. For regularization of Bayesian unfolding, 
convergence is determined by the stability of the unfolded solution for successive iterations, which is 
achieved typically by the second iteration. 

For both unfolding approaches, 
consistency of the solution is checked by backfolding, i.e. smearing the 
unfolded distribution with \Rfull\ and
comparing the result with the \DrecoilM\ distribution. 
Since regularization suppresses oscillating components of the 
solution, the backfolded and \DrecoilM\ distributions will in general not be identical.
Consistency of unfolding is imposed by requiring that the difference between the 
backfolded and \DrecoilM\ distributions in each bin be less than 3$\sigma$, based on 
\DrecoilM\ statistical errors; otherwise, the solution is rejected. 

Closure of the unfolding procedure was verified by a test in which 
the response matrix, the \Drecoil\ distribution, and the prior were generated by 
statistically independent sets 
of PYTHIA-generated events for \pp\ collisions at $\sqrts=5.02$\,TeV. The 
response matrix and 
the spectrum were generated using PYTHIA6 Perugia-11, while the prior was 
generated using PYTHIA8 tune 4C. The \Drecoil\ distribution from this test agrees 
with the input particle-level distribution to better than 5\%.

Correction for jet reconstruction efficiency is 
applied after the unfolding step by scaling the unfolded \Drecoil\  distribution 
by $1/\mathrm{eff}(\pTjetch)$. The value of $\mathrm{eff}(\pTjetch)$ is 0.96 
at $\pTjetch =15$\,\gev\ and 0.98 at $\pTjetch =60$\,\gev. 

\section{Systematic uncertainties}
\label{sect:Uncert}

The systematic uncertainties of the \Drecoil\ distribution are assessed by varying 
the components of the correction procedure. The most significant systematic 
uncertainties are due to the following:
 
\begin{itemize}
\item Regularization of unfolding: for SVD, vary $k$ by $\pm2$ relative to its 
value in the primary analysis; for Bayesian unfolding, use the first three 
iterations;  
\item Unfolding prior: generate prior distributions with PYTHIA6 and PYTHIA8; for additional variation take the difference between the priors from the two PYTHIA versions and vary them by its magnitude but with opposite sign; use the unfolded solution based on the iterative Bayesian approach as prior for SVD-based unfolding;
\item Binning of distributions: use three different choices of binning, with corresponding 
variation in spectrum limits; 
\item Calculation of $\rho$: utilize a modified procedure \cite{Chatrchyan:2012bg,Adam:2015hoa,Adam:2015xea} that accounts 
for sparse regions of the event, instead of the area-based approach (Eq.~\ref{eq:rho});
\item \cRef\ variation: use as upper limit $\cRef=1$, in which the 
reference recoil jet spectrum is not scaled. For the lower limit, double the value 
of $(1-\cRef)$ from the primary analysis, giving $\cRef=0.95$ for $\rr=0.2$ and 
$\cRef=0.90$ for $\rr=0.4$. The systematic uncertainty band corresponds to the 
largest deviation from all such variations of the unfolded spectrum, relative to 
the spectrum resulting from the \cRef\ choice of the primary analysis;
\item Tracking efficiency: vary $\pm4$\% relative to nominal value \cite{Adam:2015hoa};
\item Track momentum resolution: extract systematic uncertainty of momentum resolution from azimuthal variation of the inclusive charged-track distribution; vary \Rinstr\ accordingly. 
\end{itemize}

The correction for secondary vertex tracks due to weak decays makes a smaller 
contribution to the 
systematic uncertainty than the above sources.

There is a difference in the response matrix for different selections of EA, due 
to the different magnitude of uncorrelated background induced by such a 
selection. The correction procedure accounts for this difference. However, there 
may be a residual correlation between the EA-bias and TT-bias in the calculation of 
the response. This correlation was explored by calculating the response matrix 
with the appropriate EA-selected data, both with and without TT-bias. The 
corrected spectra resulting from the two response matrices differ by less than 
2\% for all \pTjetch, \rr, and EA-selection. This is however a check, not a 
systematic uncertainty, since the response matrix for the analysis is properly 
calculated using the TT-bias, and it does not contribute to the systematic uncertainty of the measurement.

The EA-bias induced by the TT\{6,7\} and TT\{12,50\} requirements are similar, and the 
\dpT\ distributions generated for events with the two TT requirements are likewise similar. This variation in the \dpT\ distribution generates variation of less that 1\% in the corrected spectrum, after unfolding.

Statistical fluctuations of the raw data influence the quantitative 
assessment of the systematic uncertainties arising from these sources. 
We utilize the following procedure to minimize such effects. For each source of 
uncertainty, several randomized instances of the raw $\Drecoil$ spectrum 
are generated by variation about the measured central value in each bin using a 
Gaussian distribution, with $\sigma$ equal to the uncorrelated statistical error 
in the bin. Each randomized instance is analyzed 
using (i) corrections for the primary analysis (see Sect.~\ref{sect:Corr}), and
(ii) corrections that include a systematic variation. For each randomized instance, the ratio of corrected $\Drecoil$ spectra resulting 
from (ii) and (i)  is formed. The systematic uncertainty in each \pTjetch-bin is defined as the median of the distribution of ratios obtained from all randomized instances.

\begin{table}[ht!]
\begin{center}
\begin{tabular}{|c|c|c|c|c|c|c|c|c|}
\hline
	& \multicolumn{4}{ c |}{\Drecoil\ syst. uncert. (\%)} &  \multicolumn{4}{ c |}{\Drecoil\ syst. uncert. (\%)}   \\
	& \multicolumn{4}{ c |}{ ZNA 0--20\%}                &  \multicolumn{4}{ c |}{  ZNA 50--100\% }   \\ \hline
\pTjetch\  & \multicolumn{2}{ c |}{15--20\,\gev}  &  \multicolumn{2}{ c |}{40--50\,\gev}  &   
\multicolumn{2}{ c |}{15--20\,\gev}  &  \multicolumn{2}{ c |}{40--50\,\gev} \\
\hline
\rr							& 0.2 &	0.4 & 0.2 &	0.4 & 0.2 &	0.4 & 0.2 &	0.4 \\
 \hline\hline
Unfolding algorithm         &  $< 1$ & 1.7  &  1.8   & 4.8  &   1.4   & 1.4  & 1.1                      & $< 1$ \\
Unfolding prior             & 0.5 &  0.2&  1.7   &  0.5 & 0.2  & 1.2  & 1.5 & 1.2 \\
Binning of raw spectrum     &  1.1    &  2.4 &  0.5 & 1.2  &   1.0  & 1.2 & 2.1 & 2.2 \\
$\rho$ estimator            &  0.2  &  2.7 &  0.9 & 0.2 &   0.8   &  2.8 & 2.0   & 4.4   \\
\cRef                       &  2.3   & 3.6  &  1.7 & 0.5 & 1.4 & 0.9 & 1.7 & 1.3 \\
Track reconstruction efficiency  &  4.7  & 3.3  &  9.0    &  11  &   4.8   & 4.2  & 10    &  11  \\ 
Track \pT\ resolution       &  0.6    &  0.6    &  1.0    & 1.7 &   0.6    &    0.6   & 1.0     &  1.7    \\ 
Weak decays	            &  $< 1$    &  $< 1$   &  $< 1$   &  $< 1$     &  $< 1$   &      $< 1$  &  $< 1$     &     $< 1$      \\ \hline
Cumulative                  &  5.4   &    6.3   &  9.6   &  12       &   5.4  &    5.6    &   11    &   12     \\ \hline
\end{tabular}
\end{center}
\caption{Contributions to the relative systematic uncertainty of the \Drecoil\ distribution for $\rr =0.2$ and 0.4 in EA-biased events based on ZNA.}
\label{tab:SysUncert}
\end{table}

Table~\ref{tab:SysUncert} gives representative systematic uncertainties for $\rr =0.2$ and $\rr=0.4$ in EA-biased events based on ZNA. 
The cumulative systematic uncertainty is calculated by adding contributions 
from all systematic sources in quadrature. For $\pTjetch=15$--20\,\gev, several components contribute with similar magnitude. For $\pTjetch =40$--50\,\gev, the cumulative uncertainty is due predominantly to the uncertainty in tracking efficiency. Similar uncertainties are obtained for event selection using the EA bias based on V0A. 

The systematic uncertainty of the ratio of \Drecoil\ distributions was obtained 
similarly, taking into account the correlated uncertainties of numerator and denominator.

\section{Results}
\label{sect:Results}

\begin{figure}[tbh!]
\centering
\includegraphics[width=0.49\textwidth]{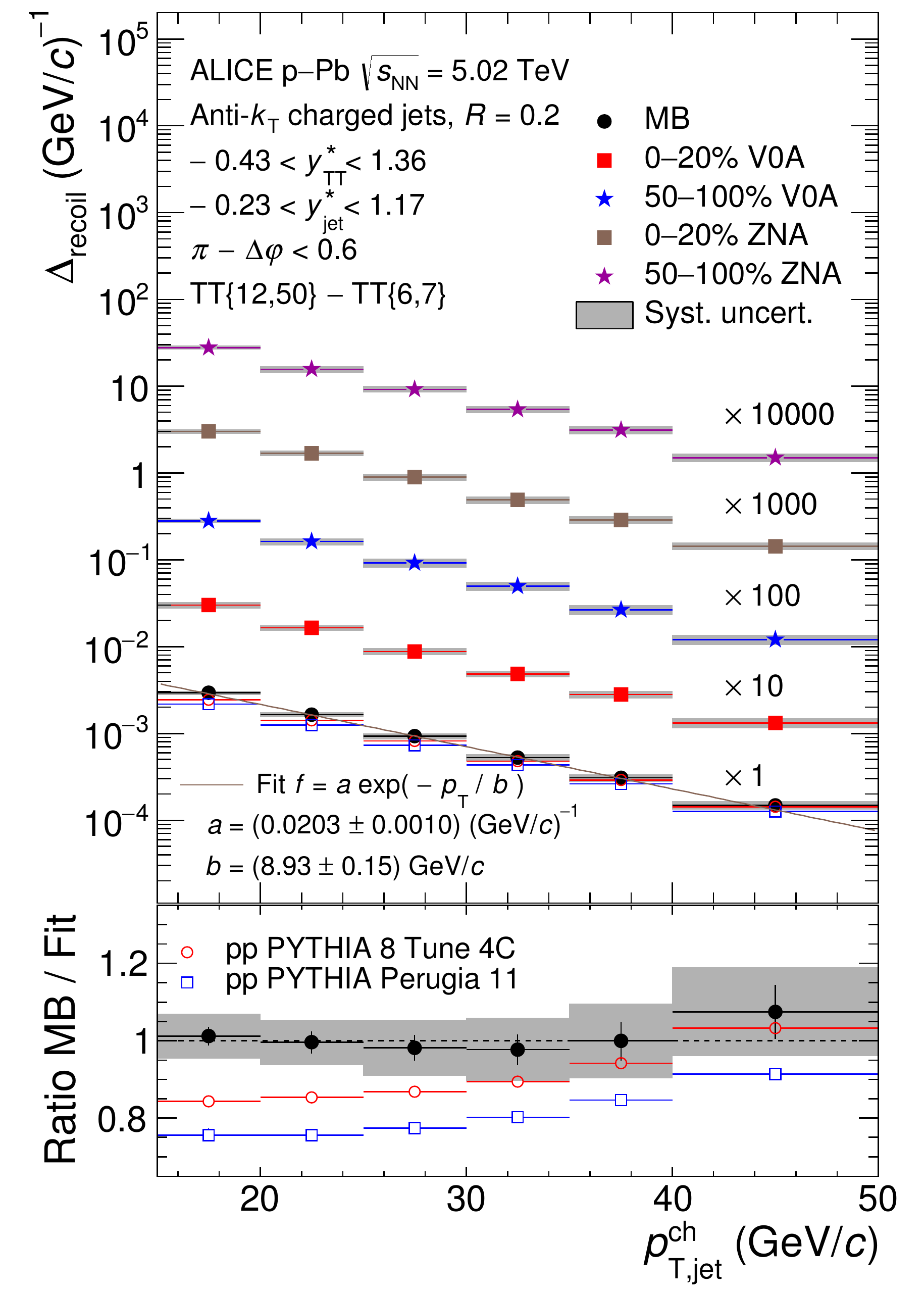}
\includegraphics[width=0.49\textwidth]{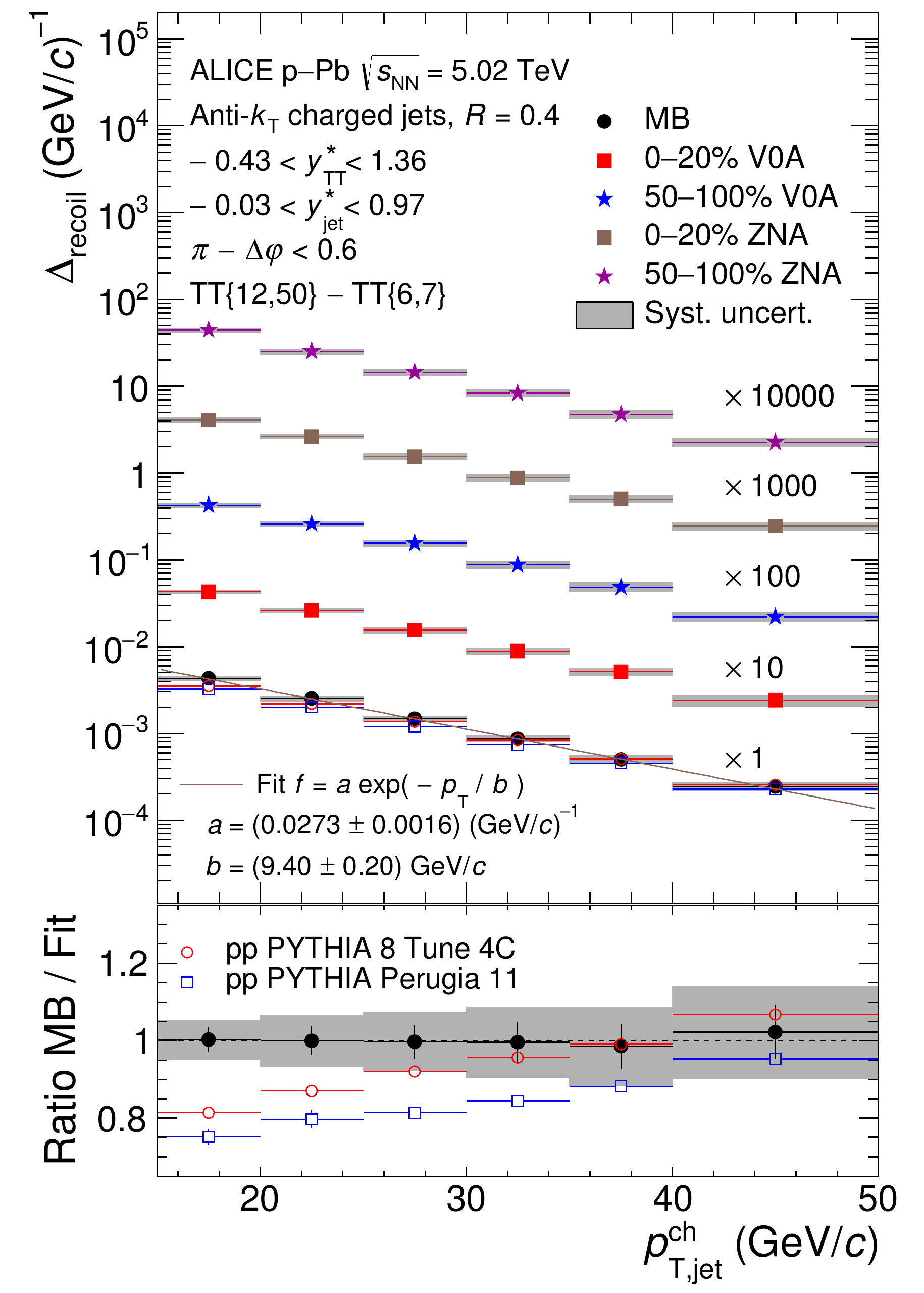}
\caption{Corrected \Drecoil\ distributions measured for \pPb\ 
collisions at 
$\sqrtsNN = 5.02$\,TeV, for the MB and EA-selected populations. The acceptance 
for TT and recoil jets in the CM frame are denoted \ystarTT\ and \ystarjet, 
respectively. Left panels: 
$\rr=0.2$; right panels: $\rr=0.4$.  
Also shown are \Drecoil\ distributions for \pp\ 
collisions at $\sqrts = 5.02$ TeV simulated by PYTHIA 6 Tune Perugia 11 and 
PYTHIA 8 
Tune 4C. The solid line in the upper panels is the fit of an exponential 
function to 
the \pPb\ distribution, with fit parameters as specified. Lower panels: ratio of 
\pPb\ MB and \pp\
distributions to the fit function. 
}
\label{fig:CorrectedSpectra}
\end{figure}

Figure~\ref{fig:CorrectedSpectra}, upper panels, show the corrected \Drecoil\
 distributions for \pPb\ collisions at 
$\sqrtsNN = 5.02$\,TeV for the MB event population and for populations selected 
by EA using ZNA and V0A, and for \pp\ collisions at $\sqrts = 5.02$ TeV  simulated by 
PYTHIA. The upper panels also show the result of a fit to the \pPb\ MB distributions by 
an exponential function, 
$a\cdot\exp\left(-\pTjetch/b\right)$. 

The PYTHIA-generated distributions for \pp\ collisions are presented only for 
comparison and are not utilized in the jet quenching analysis.
Figure~\ref{fig:CorrectedSpectra}, lower panels, show the ratio of the measured \pPb\ MB 
and PYTHIA-generated \pp\ distributions 
to the fit distribution. 
The central values of the PYTHIA-generated 
distributions for \pp\ collisions lie below those of the 
\pPb\ data, with a difference of 25\% for $\pTjetch<20$\,\gev. PYTHIA 8 tune 
4C agrees better with the \pPb\ data, notably at the highest \pTjetch\ and 
for $\rr 
=0.4$. For \pp\ collisions at $\sqrts =7$\,TeV,
PYTHIA-generated \Drecoil\ distributions have central values  that are in 
good agreement with data \cite{Adam:2015pbpb}. We note, however, that the \pTtrig\ intervals 
in the two analyses are different: this analysis uses TT\{12,50\}, whereas that 
in ref.~\cite{Adam:2015pbpb} used TT\{20,50\}. Reanalysis of the \pp\ 7\,TeV 
data with the trigger selection TT\{12,50\} shows a similar level of agreement 
with 
PYTHIA as that seen in Fig.~\ref{fig:CorrectedSpectra}~\cite{PublicNote}. We also 
note 
that at present there are significant uncertainties in the light hadron fragmentation 
functions at LHC energies~\cite{d'Enterria:2013vba, deFlorian:2014xna}, which 
may affect hadron trigger selection in the PYTHIA calculation and thereby 
contribute to the differences between PYTHIA and data seen in the figure. 

The  \Drecoil\ distributions for EA-selected event populations and for the MB 
population shown in Fig.~\ref{fig:CorrectedSpectra} are all qualitatively
similar. Measurement of the dependence of the \Drecoil\ distribution on EA selection is therefore carried out using the ratios of such distributions, denoted \REA, to maximize the sensitivity to variations with EA. 

\begin{figure}[tbh!]
\centering
\includegraphics[width=0.48\textwidth]{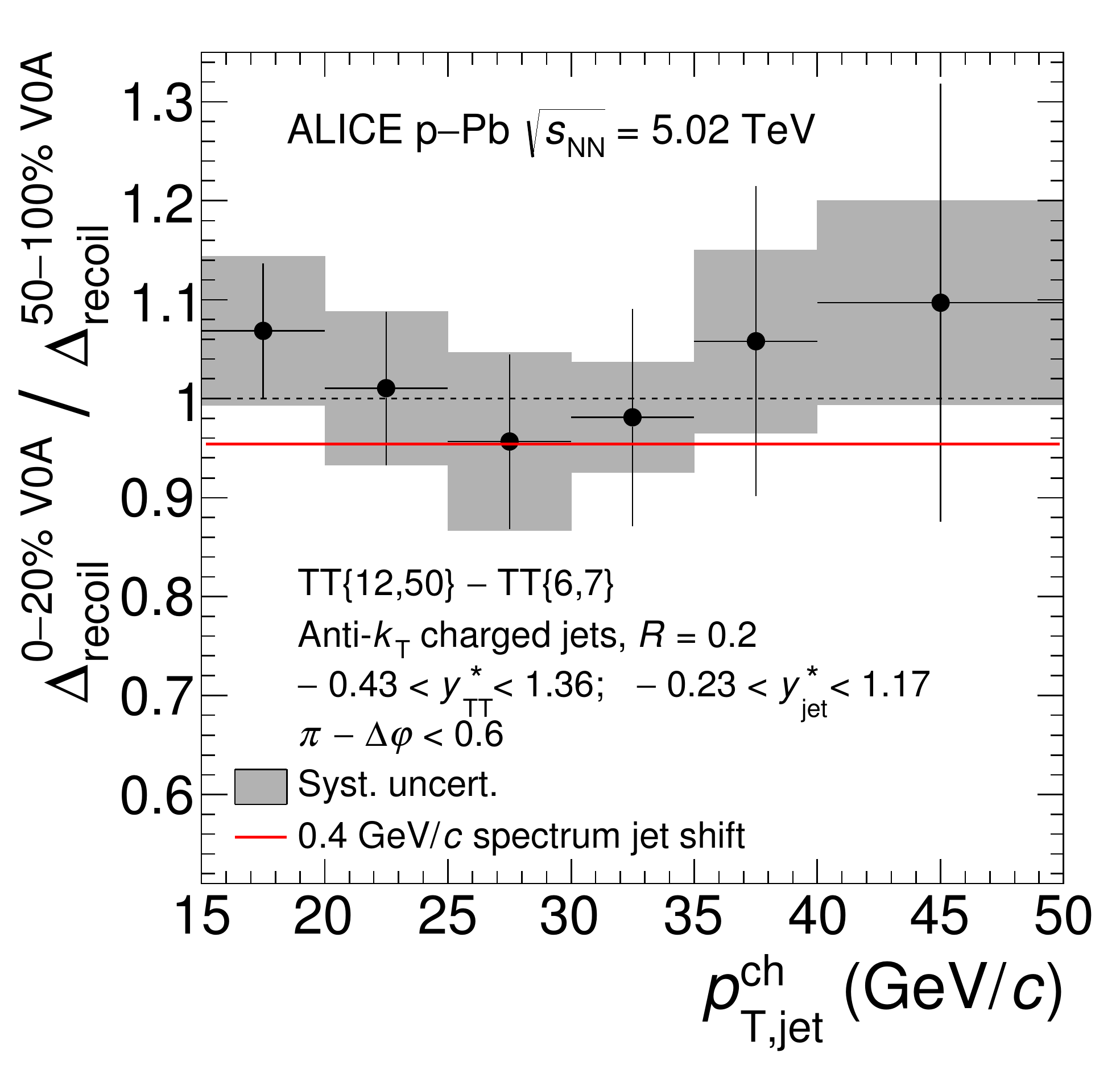}
\includegraphics[width=0.48\textwidth]{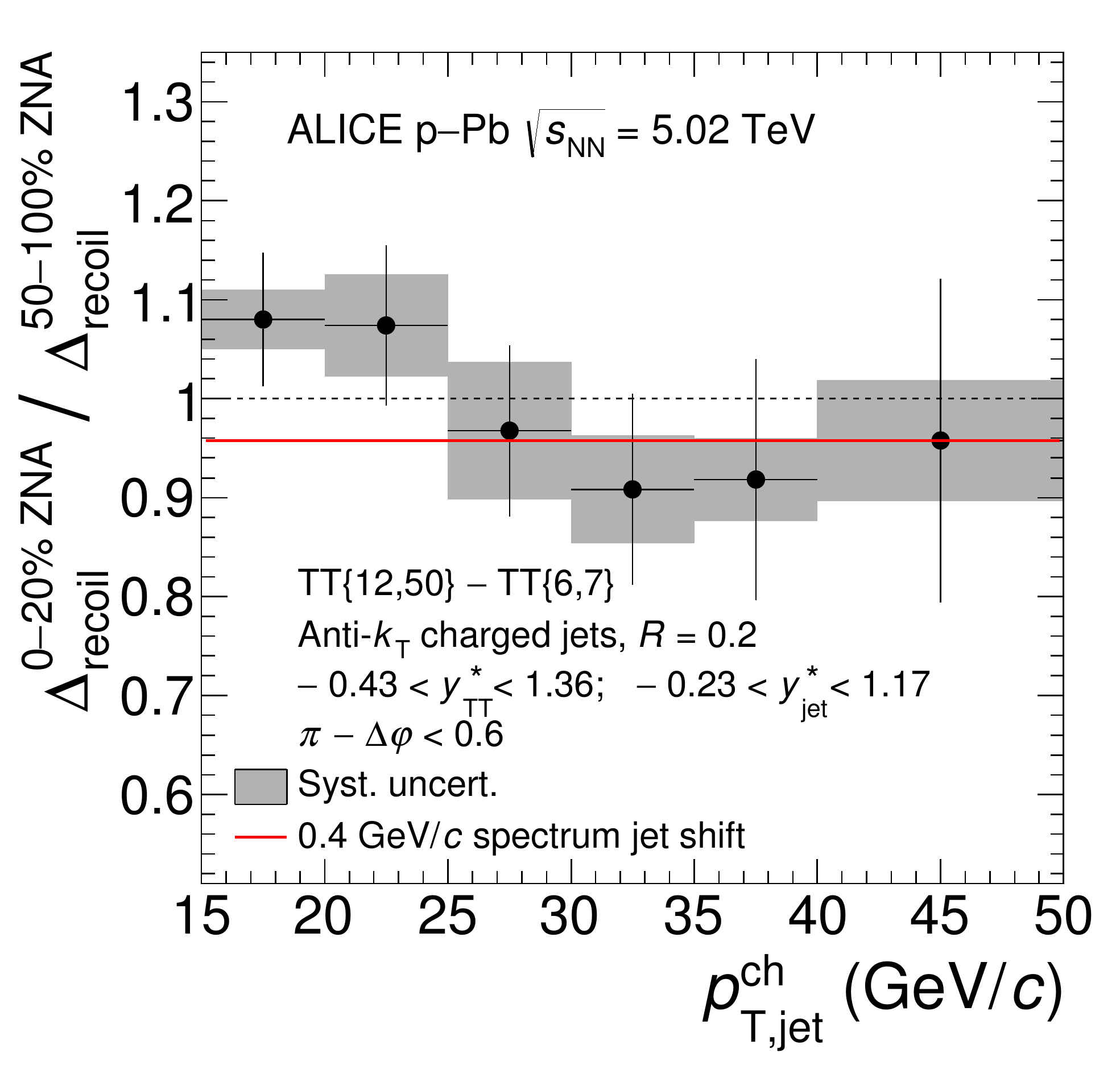}
\includegraphics[width=0.48\textwidth]{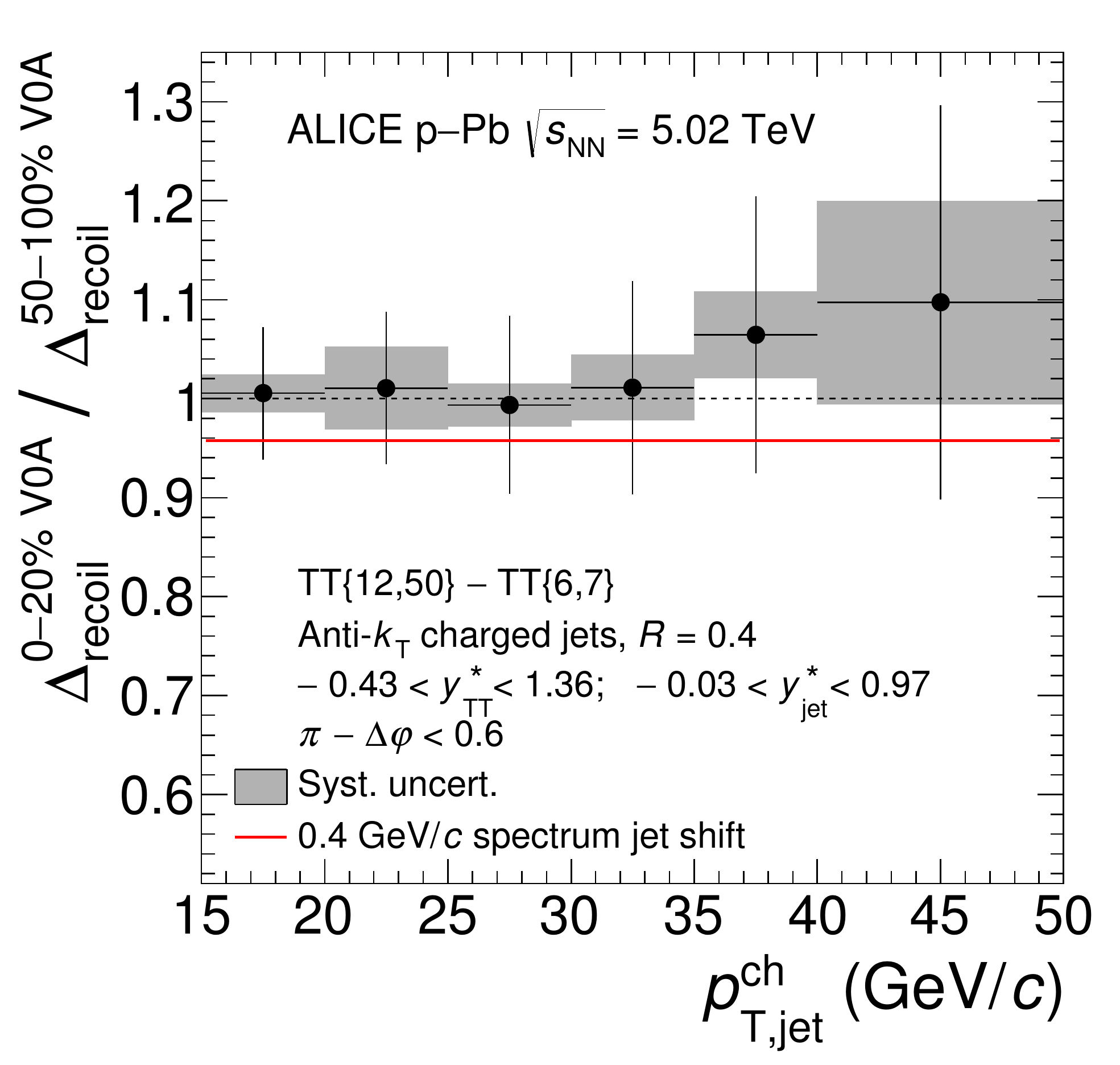}
\includegraphics[width=0.48\textwidth]{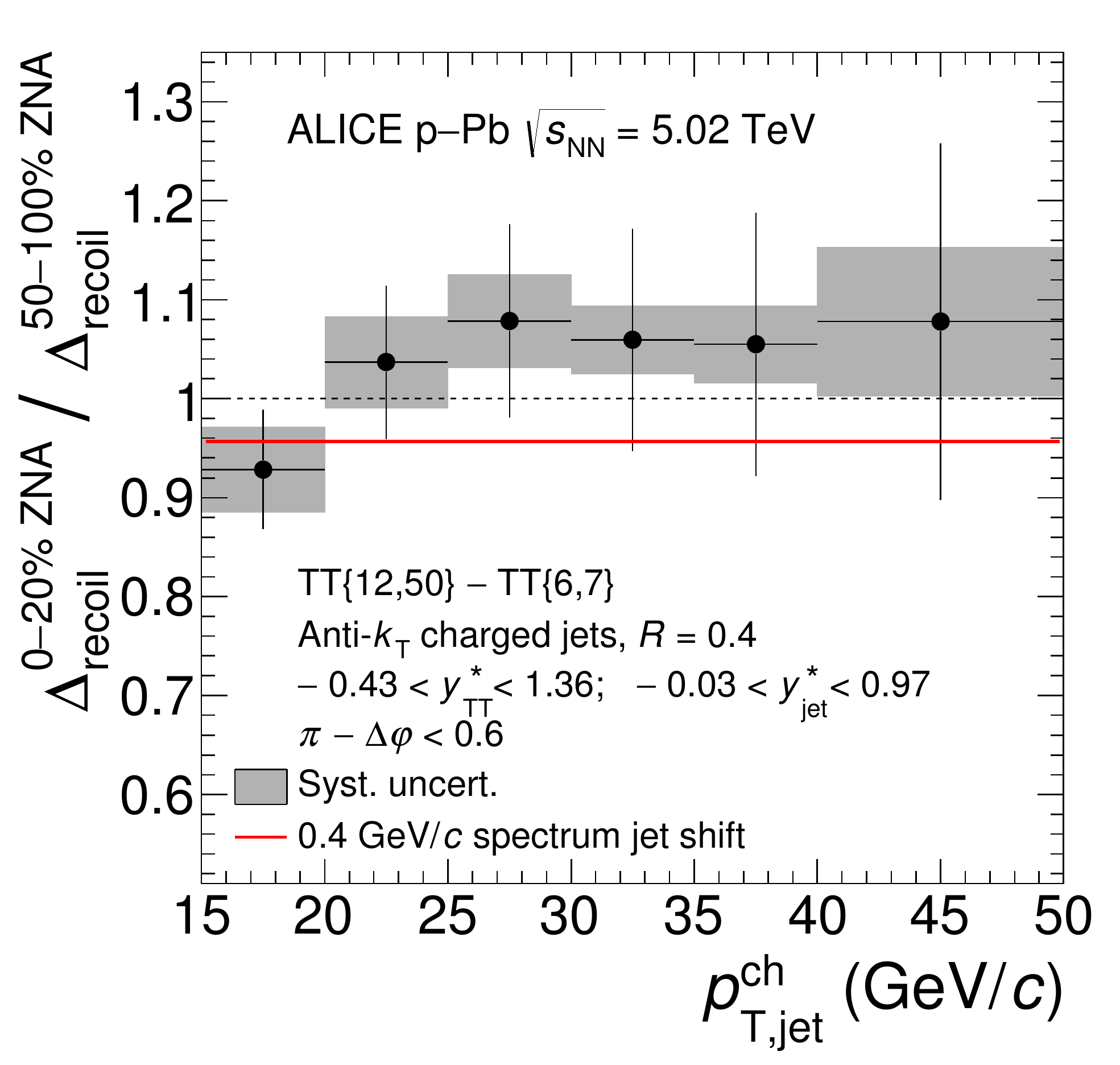}
\caption{Ratio of \Drecoil\ distributions for events with high and low EA measured in \pPb\ 
collisions at $\sqrtsNN =5.02$\,TeV. Left panels: V0A 0--20\% / 50--100\%; 
right panels: ZNA 0--20\% / 50--100\%. Upper panels: $\rr= 0.2$; 
lower panels: $\rr =0.4$. The grey boxes show the systematic uncertainty of the ratio, 
which takes into account the correlated uncertainty of numerator and denominator.
The red line indicates the ratio for a \pT-shift of the high-EA distribution of $-0.4$\,\gev. 
}
\label{fig:RatioDeltaRecoilR0204}
\end{figure}

Figure~\ref{fig:RatioDeltaRecoilR0204} shows ratios of the \Drecoil\ distributions 
for EA-selected event populations, with $\rr =0.2$ and 0.4. Since the 
numerator and denominator come from different, exclusive intervals in EA, they 
are 
statistically independent in each panel. However, some systematic uncertainties 
are 
correlated between numerator and denominator, which has been taken 
into account in the systematic uncertainty of the ratio. Note that 
the same dataset is used for $\rr =0.2$ and $\rr =0.4$, and for the ZNA and V0A 
selections, so that the results shown in the different panels are correlated. 

Jet quenching may result in transport of jet energy out of the jet cone, 
resulting in  suppression of the \Drecoil\ distribution at fixed \pTjetch. 
Under the assumptions (i) that jet 
quenching is more likely to occur in events with larger EA, and (ii) that the 
hadron trigger bias is independent of EA (Sect.~\ref{sect:Observable}), this 
effect corresponds to suppression below unity of the ratios in 
Fig.~\ref{fig:RatioDeltaRecoilR0204}. However, in all panels the ratio is 
consistent with unity within the statistical error and the 
systematic uncertainty at all \pTjetch, indicating that jet quenching effects 
are negligible relative to the uncertainties. 

These data can nevertheless provide a limit 
on the magnitude of medium-induced energy 
transport to large angles. In order to extract a limit, we parameterize the 
0--20\% and 50--100\% EA-selected \Drecoil\ distributions with the exponential 
function used in Fig.~\ref{fig:CorrectedSpectra}, and assume that the slope parameter $b$ is the same for the two distributions. We also assume that the average 
magnitude of energy transported out-of-cone is independent of \pTjetch, 
which is consistent with the observation that
the ratios \REA\ in 
Fig.~\ref{fig:RatioDeltaRecoilR0204} are independent of \pTjetch\ within 
uncertainties. The assumption that the average magnitude of out-of-cone 
radiation is independent of \pTjetch\ is likewise consistent with \Drecoil\ measurements 
in \PbPb\ collisions at 2.76 TeV~\cite{Adam:2015pbpb}. Consideration of a more 
complex dependence on \pTjetch\ is beyond the scope of this phenomenological 
study.

The ratios \REA\ are then expressed in terms of an 
average shift $\bar{s}$ in \pTjetch\ between low and high EA events, where 
$\bar{s}=-b\cdot\ln\left(\REA\right)$. Fits to \Drecoil\ for $\rr =0.4$ over the 
range $15<\pTjetch<50$\,\gev\ give 
$b=9.26\pm0.33$\,\gev\ for 50--100\% ZNA and $b=9.05\pm0.30$\,\gev\ for  
50--100\% V0A. Fits to the ratios in Fig.~\ref{fig:RatioDeltaRecoilR0204} then 
give $\bar{s}=(-0.12\pm0.35_{\rm{stat}}\pm0.03_{\rm{syst}})$\,\gev\ for 0--20\% 
ZNA, and $\bar{s}=(-0.06\pm0.34_{\rm{stat}}\pm0.02_{\rm{syst}})$\,\gev\ for  
0--20\% V0A, both of which are consistent with zero within uncertainties. 
Fits to narrower ranges in \pTjetch\ give similar results. 

These values 
are to be compared with the shift $\bar{s}=(8\pm2_{\rm{stat}})$\,\gev\ measured 
in central \PbPb\ collisions at $\sqrtsNN\ = 2.76$\,TeV for 
$\rr=0.5$~\cite{Adam:2015pbpb}, indicating significant medium-induced energy 
transport to large angles in that collision system. This comparison of 
out-of-cone energy transport in \pPb\ and \PbPb\ 
collisions supports theoretical calculations which predict much smaller jet 
quenching effects in \pPb\ relative to \PbPb\ 
collisions~\cite{Tyw:2014ppb,Chen:2015qmd}, and disfavors the 
calculation which predicts strong jet quenching in small 
systems~\cite{Zakharov:2013gya}. 

The measured value of $\bar{s}$ provides a constraint on the magnitude of 
out-of-cone energy transport due to jet quenching in \pPb\ collisions. We calculate this 
constraint as the linear sum of the central value of $\bar{s}$, the one-sided 
90\% confidence upper limit of its statistical error, and the absolute value of 
its systematic uncertainty.
For jets with $\rr =0.4$ in the 
range $15<\pTjetch<50$\,\gev, the 
medium-induced charged energy transport out of the jet cone for events with 
high V0A or high ZNA is less than 0.4\,\gev, at 90\% confidence. 
The red line in each panel of 
Fig.~\ref{fig:RatioDeltaRecoilR0204} shows the ratio for a \pT-shift of $-0.4$\,\gev\ of the high-EA distribution relative to the low-EA  distribution.

\section{Comparison to other measurements}
\label{sect:Comparison}

The EA-selected \Drecoil\ distribution ratios in 
Fig.~\ref{fig:RatioDeltaRecoilR0204} are consistent with unity in the range 
$15<\pTjetch<50$\,\gev. These distributions therefore have no significant 
dependence on EA, in agreement with inclusive jet measurements for \pPb\ 
collisions at $\sqrtsNN=5.02$\,TeV by ALICE~\cite{Adam:2016ppb}, but
in contrast to such measurements  
in \dAu\ collisions at $\sqrtsNN=200$\,GeV by PHENIX~\cite{Adare:2016dau} and in \pPb\ 
collisions at $\sqrtsNN=5.02$\,TeV by ATLAS~\cite{Aad:2015ppb}, which exhibit strong
dependence on EA. In this section we explore whether these inclusive and 
coincidence measurements can provide a consistent picture of jet quenching in asymmetric systems.  

We first note that these measurements differ in several aspects, and that their
detailed comparison requires calculations based on theoretical models of jet 
quenching that are beyond the scope of this paper.
Here we explore a more limited question, whether the inclusive 
jet measurements are also consistent with a \pT-independent limit of 
out-of-cone charged-energy transport of 
0.4 \gev. Since the ALICE inclusive jet measurement does not find an 
EA-dependence, it is consistent with such a limit by construction, without the 
need for additional calculation. We therefore focus in the rest of this section 
on comparison to the PHENIX and  ATLAS inclusive jet measurements.

To do so, we compare the effect of a \pT-independent shift of the 
inclusive spectra to the measured 
EA-dependence 
of \RCP, which is the ratio of inclusive jet spectra for event populations 
identified as ``central" and ``peripheral'', with the spectra scaled by 
$\left<\TpPb\right>$ or $\left<\TdAu\right>$. Since the inclusive spectra are 
measured with fully-reconstructed jets, 
including neutral energy, we increase the 90\% confidence limit 
for out-of-cone energy transport to the value 0.6\,\gev. Note in addition 
that the choice of percentile binning differs in the various measurements, which 
cannot be accounted for directly in the comparison; this 
difference should also be borne in mind when comparing the measurements.

The effect of the \pT-shift on jet yield depends upon the inclusive spectrum 
shape. In order to assess this effect we select a representative spectrum for each 
data set, impose a \pT-independent shift 
of $-0.6$\,\gev\ on this spectrum, and calculate \RCPstar, the ratio of 
distributions 
with and without the 
shift. Since \RCPstar\ represents a 90\% confidence 
limit, no uncertainty is assigned to it. 

\begin{table}[ht!]
\begin{center}
\begin{tabular}{|c|c|c|c|}
\hline
Collision system &  Comparison of spectra & \pTjet\ $\left(\gev\right)$ & \RCP\ or \RCPstar\ \\ \hline\hline
\multirow{4}{*}{\parbox[c]{3cm}{\dAu, \pp\ \\ $\sqrtsNN =0.2$\,TeV \cite{Adare:2016dau}}} 
& \multirow{2}{*}{\dAu\ 0--20\%/60--88\%} & 16  & $\RCP=0.71\pm0.01_{\rm{stat}}\pm0.03_{\rm{sys}}$ \\ \cline{3-4}
          &  & 32 & $\RCP=0.54\pm0.04_{\rm{stat}}\pm0.06_{\rm{sys}}$ \\ \cline{2-4}
          & \multirow{2}{*}{\pp\ w/wo $-0.6$\,\gev\ shift} & 15  &  $\RCPstar=0.79$\\ \cline{3-4}
          &                                          & 30  & $\RCPstar=0.85$ \\ \hline\hline
\multirow{4}{*}{\parbox[c]{3cm}{\pPb\ \\ $\sqrtsNN =5.02$\,TeV \\ $y^*=0.3$~\cite{Aad:2015ppb}}} 
& \multirow{2}{*}{\pPb\ 0--10\%/60--90\%} & 57  & $\RCP=1.09\pm0.02_{\rm{stat}}\pm0.03_{\rm{sys}}$ \\ \cline{3-4}
          &  & 113 & $\RCP=0.93\pm0.01_{\rm{stat}}\pm0.02_{\rm{sys}}$\\ \cline{2-4}
          & \multirow{2}{*}{\pPb\ MB w/wo $-0.6$\,\gev\ shift} & 50  &  $\RCPstar=0.95$\\ \cline{3-4}
          &                                          & 110  & $\RCPstar=0.97$ \\ \hline
\end{tabular}
\end{center}
\caption{Comparison of \RCP\ and \RCPstar\ for inclusive jet production in asymmetric collisions at RHIC and the LHC. See text for details.}
\label{tab:RCPcompare}
\end{table}

Table~\ref{tab:RCPcompare} compares \RCPstar\ to the values of EA-selected \RCP\ 
measured in asymmetric collisions at RHIC and the LHC. While some values are in 
agreement, \RCPstar\ and \RCP\ have opposite dependence on \pTjet\ for both datasets. Within the limits of this comparison, a \pTjetch-independent out-of-cone 
charged-energy transport of 0.4 \gev\ is not consistent with 
measurements of \RCP\ for inclusive jet production in EA-selected  \dAu\ 
collisions at RHIC and \pPb\ collisions 
at the LHC. 

Effects other than jet quenching in the final state can modify jet 
yields in nuclear collisions, in particular the initial-state effects of 
shadowing and energy loss in cold 
matter~\cite{Kang:2012kc,Kang:2015mta}. In addition, calculation of the Glauber 
scaling factor for inclusive measurements may be affected by fluctuations and 
dynamical correlations between a high $Q^2$ process and the soft observables 
used to tag 
EA~\cite{Abelev:2015cnt,Adare:2013nff,Maj:2014ppb,Perepelitsa:2014yta,Bzdak:2014rca,Armesto:2015kwa,Zakharov:2016zqc,Alvioli:2013vk,Alvioli:2014sba,Alvioli:2014eda,McGlinchey:2016ssj,Morsch:2017br}. 
Such non-quenching effects could be the origin of the inconsistency observed 
here in the EA-dependence of inclusive and coincidence jet measurements, since 
inclusive and coincidence observables have different sensitivity to 
initial-state effects, and Glauber scaling is required only for inclusive 
observables.


\section{Summary}
\label{sect:Summary}

We have reported measurements of the semi-inclusive distribution of charged jets 
recoiling from 
a  high-\pT\ hadron trigger in \pPb\ collisions at $\sqrtsNN =5.02$\,TeV, 
selected by event activity in forward (Pb-going) charged multiplicity and zero-degree 
neutral energy. Interpretation of this coincidence observable does not require 
the assumption that event activity is correlated with collision geometry, with the 
corresponding uncertainties of Glauber modeling. It 
provides  a new probe of jet quenching in \pPb\ collisions that is systematically 
different from quenching measurements based on inclusive jet production.

No significant difference is observed in recoil jet distributions for different 
event activity. This measurement provides a limit on 
out-of-cone energy transport due to jet quenching in \pPb\ collisions at the 
LHC of less that 0.4\,\gev\ at 90\% confidence for jet radius $\rr =0.4$. 
Comparison of this measurement to theoretical calculations favors models with 
little or no energy loss in small systems. Comparison to inclusive jet 
measurements in small systems at RHIC and LHC indicates that the inclusive jet 
yield modification observed in EA-selected populations is not consistent with jet 
quenching. Future \pPb\ measurements at the LHC, with higher statistical 
precision and greater kinematic reach, will provide more stringent limits on 
jet quenching in light systems.

\newenvironment{acknowledgement}{\relax}{\relax}
\begin{acknowledgement}
\section*{Acknowledgements}
\label{sect:Acknowledgments}
 

The ALICE Collaboration would like to thank all its engineers and technicians for their invaluable contributions to the construction of the experiment and the CERN accelerator teams for the outstanding performance of the LHC complex.
The ALICE Collaboration gratefully acknowledges the resources and support provided by all Grid centres and the Worldwide LHC Computing Grid (WLCG) collaboration.
The ALICE Collaboration acknowledges the following funding agencies for their support in building and running the ALICE detector:
A. I. Alikhanyan National Science Laboratory (Yerevan Physics Institute) Foundation (ANSL), State Committee of Science and World Federation of Scientists (WFS), Armenia;
Austrian Academy of Sciences and Nationalstiftung f\"{u}r Forschung, Technologie und Entwicklung, Austria;
Ministry of Communications and High Technologies, National Nuclear Research Center, Azerbaijan;
Conselho Nacional de Desenvolvimento Cient\'{\i}fico e Tecnol\'{o}gico (CNPq), Universidade Federal do Rio Grande do Sul (UFRGS), Financiadora de Estudos e Projetos (Finep) and Funda\c{c}\~{a}o de Amparo \`{a} Pesquisa do Estado de S\~{a}o Paulo (FAPESP), Brazil;
Ministry of Science \& Technology of China (MSTC), National Natural Science Foundation of China (NSFC) and Ministry of Education of China (MOEC) , China;
Ministry of Science, Education and Sport and Croatian Science Foundation, Croatia;
Ministry of Education, Youth and Sports of the Czech Republic, Czech Republic;
The Danish Council for Independent Research | Natural Sciences, the Carlsberg Foundation and Danish National Research Foundation (DNRF), Denmark;
Helsinki Institute of Physics (HIP), Finland;
Commissariat \`{a} l'Energie Atomique (CEA) and Institut National de Physique Nucl\'{e}aire et de Physique des Particules (IN2P3) and Centre National de la Recherche Scientifique (CNRS), France;
Bundesministerium f\"{u}r Bildung, Wissenschaft, Forschung und Technologie (BMBF) and GSI Helmholtzzentrum f\"{u}r Schwerionenforschung GmbH, Germany;
General Secretariat for Research and Technology, Ministry of Education, Research and Religions, Greece;
National Research, Development and Innovation Office, Hungary;
Department of Atomic Energy Government of India (DAE), Department of Science and Technology, Government of India (DST), University Grants Commission, Government of India (UGC) and Council of Scientific and Industrial Research (CSIR), India;
Indonesian Institute of Science, Indonesia;
Centro Fermi - Museo Storico della Fisica e Centro Studi e Ricerche Enrico Fermi and Istituto Nazionale di Fisica Nucleare (INFN), Italy;
Institute for Innovative Science and Technology , Nagasaki Institute of Applied Science (IIST), Japan Society for the Promotion of Science (JSPS) KAKENHI and Japanese Ministry of Education, Culture, Sports, Science and Technology (MEXT), Japan;
Consejo Nacional de Ciencia (CONACYT) y Tecnolog\'{i}a, through Fondo de Cooperaci\'{o}n Internacional en Ciencia y Tecnolog\'{i}a (FONCICYT) and Direcci\'{o}n General de Asuntos del Personal Academico (DGAPA), Mexico;
Nederlandse Organisatie voor Wetenschappelijk Onderzoek (NWO), Netherlands;
The Research Council of Norway, Norway;
Commission on Science and Technology for Sustainable Development in the South (COMSATS), Pakistan;
Pontificia Universidad Cat\'{o}lica del Per\'{u}, Peru;
Ministry of Science and Higher Education and National Science Centre, Poland;
Korea Institute of Science and Technology Information and National Research Foundation of Korea (NRF), Republic of Korea;
Ministry of Education and Scientific Research, Institute of Atomic Physics and Romanian National Agency for Science, Technology and Innovation, Romania;
Joint Institute for Nuclear Research (JINR), Ministry of Education and Science of the Russian Federation and National Research Centre Kurchatov Institute, Russia;
Ministry of Education, Science, Research and Sport of the Slovak Republic, Slovakia;
National Research Foundation of South Africa, South Africa;
Centro de Aplicaciones Tecnol\'{o}gicas y Desarrollo Nuclear (CEADEN), Cubaenerg\'{\i}a, Cuba and Centro de Investigaciones Energ\'{e}ticas, Medioambientales y Tecnol\'{o}gicas (CIEMAT), Spain;
Swedish Research Council (VR) and Knut \& Alice Wallenberg Foundation (KAW), Sweden;
European Organization for Nuclear Research, Switzerland;
National Science and Technology Development Agency (NSDTA), Suranaree University of Technology (SUT) and Office of the Higher Education Commission under NRU project of Thailand, Thailand;
Turkish Atomic Energy Agency (TAEK), Turkey;
National Academy of  Sciences of Ukraine, Ukraine;
Science and Technology Facilities Council (STFC), United Kingdom;
National Science Foundation of the United States of America (NSF) and United States Department of Energy, Office of Nuclear Physics (DOE NP), United States of America.

\end{acknowledgement}
 
\bibliographystyle{utphys}
\bibliography{references}

\newpage
\appendix
\section{The ALICE Collaboration}
\label{app:collab}

\begingroup
\small
\begin{flushleft}
S.~Acharya\Irefn{org136}\And 
D.~Adamov\'{a}\Irefn{org92}\And 
J.~Adolfsson\Irefn{org33}\And 
M.M.~Aggarwal\Irefn{org97}\And 
G.~Aglieri Rinella\Irefn{org34}\And 
M.~Agnello\Irefn{org30}\And 
N.~Agrawal\Irefn{org46}\And 
Z.~Ahammed\Irefn{org136}\And 
S.U.~Ahn\Irefn{org76}\And 
S.~Aiola\Irefn{org141}
% ...
\And
\renewcommand\labelenumi{\textsuperscript{\theenumi}~}

\section*{Affiliation notes}
\renewcommand\theenumi{\roman{enumi}}
\begin{Authlist}
\item \Adef{org*}Deceased
\item \Adef{orgI}Dipartimento DET del Politecnico di Torino, Turin, Italy
\item \Adef{orgII}M.V. Lomonosov Moscow State University, D.V. Skobeltsyn Institute of Nuclear, Physics, Moscow, Russia
\item \Adef{orgIII}Department of Applied Physics, Aligarh Muslim University, Aligarh, India
\item \Adef{orgIV}Institute of Theoretical Physics, University of Wroclaw, Poland
\end{Authlist}

\section*{Collaboration Institutes}
\renewcommand\theenumi{\arabic{enumi}~}
\begin{Authlist}
\item \Idef{org1}A.I. Alikhanyan National Science Laboratory (Yerevan Physics Institute) Foundation, Yerevan, Armenia
\item \Idef{org2}Benem\'{e}rita Universidad Aut\'{o}noma de Puebla, Puebla, Mexico
\item \Idef{org3}Bogolyubov Institute for Theoretical Physics, Kiev, Ukraine
\item \Idef{org4}Bose Institute, Department of Physics  and Centre for Astroparticle Physics and Space Science (CAPSS), Kolkata, India
\item \Idef{org5}Budker Institute for Nuclear Physics, Novosibirsk, Russia
\item \Idef{org6}California Polytechnic State University, San Luis Obispo, California, United States
\item \Idef{org7}Central China Normal University, Wuhan, China
\item \Idef{org8}Centre de Calcul de l'IN2P3, Villeurbanne, Lyon, France
\item \Idef{org9}Centro de Aplicaciones Tecnol\'{o}gicas y Desarrollo Nuclear (CEADEN), Havana, Cuba
\item \Idef{org10}Centro de Investigaci\'{o}n y de Estudios Avanzados (CINVESTAV), Mexico City and M\'{e}rida, Mexico
\item \Idef{org11}Centro Fermi - Museo Storico della Fisica e Centro Studi e Ricerche ``Enrico Fermi', Rome, Italy
\item \Idef{org12}Chicago State University, Chicago, Illinois, United States
\item \Idef{org13}China Institute of Atomic Energy, Beijing, China
\item \Idef{org14}Chonbuk National University, Jeonju, Republic of Korea
\item \Idef{org15}COMSATS Institute of Information Technology (CIIT), Islamabad, Pakistan
\item \Idef{org16}Department of Physics, Aligarh Muslim University, Aligarh, India
\item \Idef{org17}Department of Physics, Ohio State University, Columbus, Ohio, United States
\item \Idef{org18}Department of Physics, Pusan National University, Pusan, Republic of Korea
\item \Idef{org19}Department of Physics, Sejong University, Seoul, Republic of Korea
\item \Idef{org20}Department of Physics, University of Oslo, Oslo, Norway
\item \Idef{org21}Department of Physics and Technology, University of Bergen, Bergen, Norway
\item \Idef{org22}Dipartimento di Fisica dell'Universit\`{a} 'La Sapienza' and Sezione INFN, Rome, Italy
\item \Idef{org23}Dipartimento di Fisica dell'Universit\`{a} and Sezione INFN, Cagliari, Italy
\item \Idef{org24}Dipartimento di Fisica dell'Universit\`{a} and Sezione INFN, Trieste, Italy
\item \Idef{org25}Dipartimento di Fisica dell'Universit\`{a} and Sezione INFN, Turin, Italy
\item \Idef{org26}Dipartimento di Fisica e Astronomia dell'Universit\`{a} and Sezione INFN, Bologna, Italy
\item \Idef{org27}Dipartimento di Fisica e Astronomia dell'Universit\`{a} and Sezione INFN, Catania, Italy
\item \Idef{org28}Dipartimento di Fisica e Astronomia dell'Universit\`{a} and Sezione INFN, Padova, Italy
\item \Idef{org29}Dipartimento di Fisica `E.R.~Caianiello' dell'Universit\`{a} and Gruppo Collegato INFN, Salerno, Italy
\item \Idef{org30}Dipartimento DISAT del Politecnico and Sezione INFN, Turin, Italy
\item \Idef{org31}Dipartimento di Scienze e Innovazione Tecnologica dell'Universit\`{a} del Piemonte Orientale and INFN Sezione di Torino, Alessandria, Italy
\item \Idef{org32}Dipartimento Interateneo di Fisica `M.~Merlin' and Sezione INFN, Bari, Italy
\item \Idef{org33}Division of Experimental High Energy Physics, University of Lund, Lund, Sweden
\item \Idef{org34}European Organization for Nuclear Research (CERN), Geneva, Switzerland
\item \Idef{org35}Faculty of Engineering, Bergen University College, Bergen, Norway
\item \Idef{org36}Faculty of Mathematics, Physics and Informatics, Comenius University, Bratislava, Slovakia
\item \Idef{org37}Faculty of Nuclear Sciences and Physical Engineering, Czech Technical University in Prague, Prague, Czech Republic
\item \Idef{org38}Faculty of Science, P.J.~\v{S}af\'{a}rik University, Ko\v{s}ice, Slovakia
\item \Idef{org39}Faculty of Technology, Buskerud and Vestfold University College, Tonsberg, Norway
\item \Idef{org40}Frankfurt Institute for Advanced Studies, Johann Wolfgang Goethe-Universit\"{a}t Frankfurt, Frankfurt, Germany
\item \Idef{org41}Gangneung-Wonju National University, Gangneung, Republic of Korea
\item \Idef{org42}Gauhati University, Department of Physics, Guwahati, India
\item \Idef{org43}Helmholtz-Institut f\"{u}r Strahlen- und Kernphysik, Rheinische Friedrich-Wilhelms-Universit\"{a}t Bonn, Bonn, Germany
\item \Idef{org44}Helsinki Institute of Physics (HIP), Helsinki, Finland
\item \Idef{org45}Hiroshima University, Hiroshima, Japan
\item \Idef{org46}Indian Institute of Technology Bombay (IIT), Mumbai, India
\item \Idef{org47}Indian Institute of Technology Indore, Indore, India
\item \Idef{org48}Indonesian Institute of Sciences, Jakarta, Indonesia
\item \Idef{org49}INFN, Laboratori Nazionali di Frascati, Frascati, Italy
\item \Idef{org50}INFN, Sezione di Bari, Bari, Italy
\item \Idef{org51}INFN, Sezione di Bologna, Bologna, Italy
\item \Idef{org52}INFN, Sezione di Cagliari, Cagliari, Italy
\item \Idef{org53}INFN, Sezione di Catania, Catania, Italy
\item \Idef{org54}INFN, Sezione di Padova, Padova, Italy
\item \Idef{org55}INFN, Sezione di Roma, Rome, Italy
\item \Idef{org56}INFN, Sezione di Torino, Turin, Italy
\item \Idef{org57}INFN, Sezione di Trieste, Trieste, Italy
\item \Idef{org58}Inha University, Incheon, Republic of Korea
\item \Idef{org59}Institut de Physique Nucl\'eaire d'Orsay (IPNO), Universit\'e Paris-Sud, CNRS-IN2P3, Orsay, France
\item \Idef{org60}Institute for Nuclear Research, Academy of Sciences, Moscow, Russia
\item \Idef{org61}Institute for Subatomic Physics of Utrecht University, Utrecht, Netherlands
\item \Idef{org62}Institute for Theoretical and Experimental Physics, Moscow, Russia
\item \Idef{org63}Institute of Experimental Physics, Slovak Academy of Sciences, Ko\v{s}ice, Slovakia
\item \Idef{org64}Institute of Physics, Academy of Sciences of the Czech Republic, Prague, Czech Republic
\item \Idef{org65}Institute of Physics, Bhubaneswar, India
\item \Idef{org66}Institute of Space Science (ISS), Bucharest, Romania
\item \Idef{org67}Institut f\"{u}r Informatik, Johann Wolfgang Goethe-Universit\"{a}t Frankfurt, Frankfurt, Germany
\item \Idef{org68}Institut f\"{u}r Kernphysik, Johann Wolfgang Goethe-Universit\"{a}t Frankfurt, Frankfurt, Germany
\item \Idef{org69}Instituto de Ciencias Nucleares, Universidad Nacional Aut\'{o}noma de M\'{e}xico, Mexico City, Mexico
\item \Idef{org70}Instituto de F\'{i}sica, Universidade Federal do Rio Grande do Sul (UFRGS), Porto Alegre, Brazil
\item \Idef{org71}Instituto de F\'{\i}sica, Universidad Nacional Aut\'{o}noma de M\'{e}xico, Mexico City, Mexico
\item \Idef{org72}IRFU, CEA, Universit\'{e} Paris-Saclay, Saclay, France
\item \Idef{org73}iThemba LABS, National Research Foundation, Somerset West, South Africa
\item \Idef{org74}Joint Institute for Nuclear Research (JINR), Dubna, Russia
\item \Idef{org75}Konkuk University, Seoul, Republic of Korea
\item \Idef{org76}Korea Institute of Science and Technology Information, Daejeon, Republic of Korea
\item \Idef{org77}KTO Karatay University, Konya, Turkey
\item \Idef{org78}Laboratoire de Physique Subatomique et de Cosmologie, Universit\'{e} Grenoble-Alpes, CNRS-IN2P3, Grenoble, France
\item \Idef{org79}Lawrence Berkeley National Laboratory, Berkeley, California, United States
\item \Idef{org80}Nagasaki Institute of Applied Science, Nagasaki, Japan
\item \Idef{org81}Nara Women{'}s University (NWU), Nara, Japan
\item \Idef{org82}National and Kapodistrian University of Athens, School of Science, Department of Physics , Athens, Greece
\item \Idef{org83}National Centre for Nuclear Studies, Warsaw, Poland
\item \Idef{org84}National Institute for Physics and Nuclear Engineering, Bucharest, Romania
\item \Idef{org85}National Institute of Science Education and Research, HBNI, Jatni, India
\item \Idef{org86}National Nuclear Research Center, Baku, Azerbaijan
\item \Idef{org87}National Research Centre Kurchatov Institute, Moscow, Russia
\item \Idef{org88}Niels Bohr Institute, University of Copenhagen, Copenhagen, Denmark
\item \Idef{org89}Nikhef, Nationaal instituut voor subatomaire fysica, Amsterdam, Netherlands
\item \Idef{org90}NRNU Moscow Engineering Physics Institute, Moscow, Russia
\item \Idef{org91}Nuclear Physics Group, STFC Daresbury Laboratory, Daresbury, United Kingdom
\item \Idef{org92}Nuclear Physics Institute, Academy of Sciences of the Czech Republic, \v{R}e\v{z} u Prahy, Czech Republic
\item \Idef{org93}Oak Ridge National Laboratory, Oak Ridge, Tennessee, United States
\item \Idef{org94}Petersburg Nuclear Physics Institute, Gatchina, Russia
\item \Idef{org95}Physics Department, Creighton University, Omaha, Nebraska, United States
\item \Idef{org96}Physics department, Faculty of science, University of Zagreb, Zagreb, Croatia
\item \Idef{org97}Physics Department, Panjab University, Chandigarh, India
\item \Idef{org98}Physics Department, University of Cape Town, Cape Town, South Africa
\item \Idef{org99}Physics Department, University of Jammu, Jammu, India
\item \Idef{org100}Physics Department, University of Rajasthan, Jaipur, India
\item \Idef{org101}Physikalisches Institut, Eberhard Karls Universit\"{a}t T\"{u}bingen, T\"{u}bingen, Germany
\item \Idef{org102}Physikalisches Institut, Ruprecht-Karls-Universit\"{a}t Heidelberg, Heidelberg, Germany
\item \Idef{org103}Physik Department, Technische Universit\"{a}t M\"{u}nchen, Munich, Germany
\item \Idef{org104}Research Division and ExtreMe Matter Institute EMMI, GSI Helmholtzzentrum f\"ur Schwerionenforschung GmbH, Darmstadt, Germany
\item \Idef{org105}Rudjer Bo\v{s}kovi\'{c} Institute, Zagreb, Croatia
\item \Idef{org106}Russian Federal Nuclear Center (VNIIEF), Sarov, Russia
\item \Idef{org107}Saha Institute of Nuclear Physics, Kolkata, India
\item \Idef{org108}School of Physics and Astronomy, University of Birmingham, Birmingham, United Kingdom
\item \Idef{org109}Secci\'{o}n F\'{\i}sica, Departamento de Ciencias, Pontificia Universidad Cat\'{o}lica del Per\'{u}, Lima, Peru
\item \Idef{org110}SSC IHEP of NRC Kurchatov institute, Protvino, Russia
\item \Idef{org111}Stefan Meyer Institut f\"{u}r Subatomare Physik (SMI), Vienna, Austria
\item \Idef{org112}SUBATECH, IMT Atlantique, Universit\'{e} de Nantes, CNRS-IN2P3, Nantes, France
\item \Idef{org113}Suranaree University of Technology, Nakhon Ratchasima, Thailand
\item \Idef{org114}Technical University of Ko\v{s}ice, Ko\v{s}ice, Slovakia
\item \Idef{org115}Technical University of Split FESB, Split, Croatia
\item \Idef{org116}Technische Universit\"{a}t M\"{u}nchen, Excellence Cluster 'Universe', Munich, Germany
\item \Idef{org117}The Henryk Niewodniczanski Institute of Nuclear Physics, Polish Academy of Sciences, Cracow, Poland
\item \Idef{org118}The University of Texas at Austin, Austin, Texas, United States
\item \Idef{org119}Universidad Aut\'{o}noma de Sinaloa, Culiac\'{a}n, Mexico
\item \Idef{org120}Universidade de S\~{a}o Paulo (USP), S\~{a}o Paulo, Brazil
\item \Idef{org121}Universidade Estadual de Campinas (UNICAMP), Campinas, Brazil
\item \Idef{org122}Universidade Federal do ABC, Santo Andre, Brazil
\item \Idef{org123}University of Houston, Houston, Texas, United States
\item \Idef{org124}University of Jyv\"{a}skyl\"{a}, Jyv\"{a}skyl\"{a}, Finland
\item \Idef{org125}University of Liverpool, Liverpool, United Kingdom
\item \Idef{org126}University of Tennessee, Knoxville, Tennessee, United States
\item \Idef{org127}University of the Witwatersrand, Johannesburg, South Africa
\item \Idef{org128}University of Tokyo, Tokyo, Japan
\item \Idef{org129}University of Tsukuba, Tsukuba, Japan
\item \Idef{org130}Universit\'{e} Clermont Auvergne, CNRS/IN2P3, LPC, Clermont-Ferrand, France
\item \Idef{org131}Universit\'{e} de Lyon, Universit\'{e} Lyon 1, CNRS/IN2P3, IPN-Lyon, Villeurbanne, Lyon, France
\item \Idef{org132}Universit\'{e} de Strasbourg, CNRS, IPHC UMR 7178, F-67000 Strasbourg, France, Strasbourg, France
\item \Idef{org133}Universit\`{a} degli Studi di Pavia, Pavia, Italy
\item \Idef{org134}Universit\`{a} di Brescia, Brescia, Italy
\item \Idef{org135}V.~Fock Institute for Physics, St. Petersburg State University, St. Petersburg, Russia
\item \Idef{org136}Variable Energy Cyclotron Centre, Kolkata, India
\item \Idef{org137}Warsaw University of Technology, Warsaw, Poland
\item \Idef{org138}Wayne State University, Detroit, Michigan, United States
\item \Idef{org139}Westf\"{a}lische Wilhelms-Universit\"{a}t M\"{u}nster, Institut f\"{u}r Kernphysik, M\"{u}nster, Germany
\item \Idef{org140}Wigner Research Centre for Physics, Hungarian Academy of Sciences, Budapest, Hungary
\item \Idef{org141}Yale University, New Haven, Connecticut, United States
\item \Idef{org142}Yonsei University, Seoul, Republic of Korea
\item \Idef{org143}Zentrum f\"{u}r Technologietransfer und Telekommunikation (ZTT), Fachhochschule Worms, Worms, Germany
\end{Authlist}
\endgroup

\end{document}